\documentclass[11pt, a4paper]{article}

\usepackage[utf8]{inputenc}
\usepackage[T1]{fontenc}
\usepackage[margin=2.5cm]{geometry}
\usepackage{parskip}

\usepackage{amsmath, amssymb, amsfonts}
\usepackage{mathtools}
\usepackage{bm}
\usepackage{graphicx}
\usepackage[authoryear]{natbib}
\usepackage{url}
\usepackage{hyperref}
\usepackage{xcolor}
\usepackage{float}
\usepackage{booktabs}
\usepackage{microtype}
\usepackage{subcaption}
\usepackage{wrapfig}
\usepackage[ruled, vlined, linesnumbered]{algorithm2e}
\usepackage[most]{tcolorbox}
\usepackage{tikz}
\usetikzlibrary{patterns, positioning, decorations.pathreplacing}
\usepackage{lastpage}
\usepackage{fancyhdr}
\usepackage{fancyhdr}

\title
{Bridging the Reality Gap in Limit Order Book Simulation\footnote{Mathieu Rosenbaum and Saad Souilmi gratefully acknowledge support from the ILB Chair \textit{Artificial Intelligence and Quantitative Methods for Finance} at University Paris Dauphine-PSL. The authors thank Bruno Durin, Pierre Laffitte, Mazigh Lahiani and Charles-Albert Lehalle for inspiring discussions.}}

\author
{Patrick Noble\footnote{Jump Trading, 
\texttt{patricklnoble@gmail.com
}} \and Mathieu Rosenbaum\footnote{Universit\'e Paris Dauphine-PSL, \texttt{mathieu.rosenbaum@dauphine.psl.eu}} \and Saad Souilmi\footnote{\'Ecole Polytechnique,   \texttt{saad.souilmi@polytechnique.edu}}}

\date{\today}
\begin{document}
\maketitle

\begin{abstract}
We introduce a practical, interactive simulator of the limit order book for large-tick assets, designed to produce realistic execution, costs, and P\&L. The book state is projected onto a tractable representation based on spread and volume imbalance, enabling robust estimation from market data. Event timing is calibrated to reproduce the fine-scale temporal structure of real markets, revealing a pronounced mode at exchange round-trip latency consistent with simultaneous reactions and latency races among participants. We further incorporate a feedback mechanism that accumulates signed trade flow through a power-law decay kernel, reproducing both concave market impact during execution and partial post-trade reversion. Across several stocks and strategy case studies, the simulator yields realistic behavior where profitability becomes highly sensitive to execution parameters. We present the approach as a practical recipe: project, estimate, validate, adapt, for building realistic limit order book simulations.
\end{abstract}


\section{Introduction}
\label{sec:intro}

Limit order book simulation is a fundamental problem for electronic market participants. Market makers rely on simulators to stress-test quoting strategies before deployment, while proprietary trading firms use them to evaluate signal-driven strategies. Brokerages, facing fiduciary obligations to their clients, depend on simulation tools to benchmark execution algorithms, estimate transaction costs, and demonstrate to regulators that their routing decisions are sound. Despite these diverse use cases, the core requirement is the same: the simulator must reproduce the statistical properties of the real market with sufficient fidelity for conclusions drawn in simulation to carry over to live trading. Crucially, the simulator must also be interactive: users must be able to place orders, receive fills, and observe how the market responds to their own activity.

Early limit order book models relied on zero-intelligence
agents~\citep{smith2003statistical, cont2013markovian}, where orders
arrive at constant rates independent of the book state. The
queue-reactive (QR) model of \citet{huang2015simulating} extended this
framework by making event intensities depend on the current state of
the book. It models the limit order book as a continuous-time Markov
jump process in which the rate of each event, that is limit order,
cancellation, or market order, is conditioned on observable queue
sizes. The model is data-driven, interpretable, and reproduces many
market statistics well. Its Markov structure makes estimation
straightforward: one needs only the empirical transition rates
conditioned on the observable state.

But the Markov assumption comes at a cost. A continuous-time Markov process
generates exponentially distributed inter-event times by construction,
regardless of how sophisticated the state conditioning is. In practice,
the distribution of inter-event times in equity markets is far from
exponential. \citet{aquilina2022quantifying} document that latency
races, that is bursts of near-simultaneous orders triggered by a common
signal, occur roughly once per minute per symbol and account for
approximately 20\% of trading volume on major exchanges. The QR model,
by treating each event as Poisson arrival, averages out
this clustering entirely. It struggles to model these sudden bursts of activity.

The memoryless assumption has a second consequence: the model has no proper
mechanism for market impact of aggressive orders beyond their effect on queues. In the QR model, a metaorder does impact the price during
execution as child orders modify queue sizes in the book. Passive impact notably is well reproduced. However, after completion of the metaorder, the book evolves as if nothing happened.
For strategy evaluation, the central use case, this is a severe issue. A
simulator that ignores the user's footprint overstates profits and
understates risk.

We retain the QR engine and extend it with three modifications that turn
it into a practical, interactive simulator:
\begin{enumerate}
  \item We project the book onto a simple state based on volume
    imbalance and spread, replacing the per-queue conditioning of the
    original model where most state configurations are too rare to
    estimate reliably.
  \item We replace the exponential inter-event time assumption with a
    flexible distribution, revealing a characteristic clustering of
    events at the exchange round-trip latency and opening the door to
    modelling competitive fill dynamics.
  \item We introduce a feedback mechanism that reproduces key aspects of market impact such as the concave rise during execution and partial reversion after. The same
    mechanism can be used to inject custom trading  signals into the book
    dynamics.
\end{enumerate}

We frame our contribution as a \emph{recipe} rather than a rigid model:
project the book onto a state $\Phi(\text{LOB})$, estimate conditional
distributions from data, and validate by comparing simulated and observed
statistics. The recipe adapts to the asset and the practitioner decides what are the key features to be reproduced.

This model is designed with large-tick assets in mind, where the tick
size is significant relative to the price of the underlying. We test it
on four constituents of the S\&P~500 that trade around the \$30
mark: \texttt{INTC}, \texttt{VZ}, \texttt{T} and \texttt{PFE}, using
 data sourced from \href{https://databento.com/}{Databento}.
Throughout the paper, we illustrate results using \texttt{PFE};
equivalent figures for the other tickers can be found in the appendix.\footnote{Code available on \href{https://github.com/SaadSouilmi/Queue-Reactive}{GitHub}.}

The remainder of the paper is organised as follows.
Section~\ref{sec:framework} presents the queue-reactive framework, our
modifications to its state space, event types, and estimation
procedure, and validates the baseline model against empirical data.
Section~\ref{sec:races} replaces the exponential inter-event time
assumption with a custom distribution, revealing an exchange latency
structure that we use to model competitive fill dynamics.
Section~\ref{sec:impact} introduces a market impact feedback
mechanism that accumulates signed trade flow through a power-law decay
kernel and biases subsequent trades toward mean reversion.
Section~\ref{sec:trading} demonstrates the practical relevance
through two case studies, a mid-frequency signal-based strategy and a
high-frequency imbalance strategy.

\section{Baseline model: A simplified QR framework with random volumes}
\label{sec:framework}
\subsection{The original QR model}
The queue-reactive (QR) model of \citet{huang2015simulating} first represents the
limit order book as a $2K$-dimensional vector of queue sizes
$X(t) = (q_{-K}(t), \ldots, q_{-1}(t), q_1(t), \ldots, q_K(t))$
centered on a fixed reference price $p_{\text{ref}}$. Then a stochastic mechanism is introduced to generate moves for $p_{\text{ref}}$ (we refer the reader to \citet{huang2015simulating} for
details). Three event types can modify the book at the various limits: limit orders
(insertions), cancellations and market orders. The key modelling
assumption is that the intensity of arrival of each event depends on the current
state of the book, hence \emph{queue-reactive}.
In its most general form, the intensity $\lambda^e$ of arrival of each event $e$ (limit order, market order or cancellation at a given limit) is a function of the
entire book state:
$\lambda^e = \lambda^e(q_{-K}, \ldots, q_{-1}, q_1, \ldots, q_K)$.

In the simplest version of the QR model, each queue
  is treated independently: $q_i$ has its own arrival and cancellation
  rates $\lambda^e(q_i)$. Estimation is then straightforward, one only
  needs enough observations per queue size. However, this independence
  assumption ignores the joint dynamics: for example, in practice, the behaviour of the best ask
  queue clearly depends on the state of the best bid, and vice versa.
  The original paper addresses this by conditioning on both best queues
  $\lambda^e(q_1, q_{-1})$, but this requires aggressive binning of
  queue sizes to accumulate enough observations per bin.

  We believe that the volume imbalance at the best level already
  captures most of the bid-ask dependency that matters. It is well established
  that order book imbalance is where most of the book's short-term
  predictiveness lies~\citep{cont2014price, stoikov2018micro, pulido2026understanding}. We therefore project the book onto imbalance and spread, encoding
  the cross-queue interaction in a single scalar while keeping
  estimation tractable.

\subsection{Order book representation}
  \label{sec:order_book_repr}

  We depart from the original QR convention of indexing queues relative
  to a fixed reference price $p_{\text{ref}}$. Instead, $q_{-1}$ always
  denotes the best bid and $q_1$ always denotes the best ask, regardless
  of the current spread that we denote by $n$. Deeper queues $q_{\pm i}$ for $i \geq 2$ sit
  $i - 1$ ticks behind the best on the bid (--) or ask (+) side.

  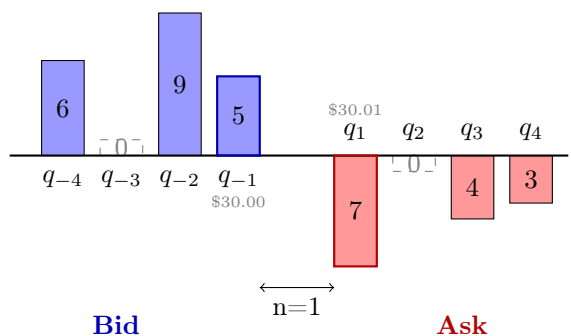
\begin{figure}[H]
  \centering
  \begin{tikzpicture}[scale=0.7]
      \foreach \x/\h/\label in {-4/6/q_{-4}, -3/0/q_{-3}, -2/9/q_{-2}, -1/5/q_{-1}} {
          \ifnum\h>0
              \fill[blue!40] (\x*1.1, 0) rectangle (\x*1.1+0.8, \h*0.3);
              \draw (\x*1.1, 0) rectangle (\x*1.1+0.8, \h*0.3);
              \node at (\x*1.1+0.4, \h*0.15) {\small\h};
          \else
              \draw[dashed, gray] (\x*1.1, 0) rectangle (\x*1.1+0.8, 0.3);
              \node[gray] at (\x*1.1+0.4, 0.15) {\small 0};
          \fi
          \node[below, font=\small] at (\x*1.1+0.4, -0.1) {$\label$};
      }

      \foreach \x/\h/\label in {1/7/q_1, 2/0/q_2, 3/4/q_3, 4/3/q_4} {
          \ifnum\h>0
              \fill[red!40] (\x*1.1, 0) rectangle (\x*1.1+0.8, -\h*0.3);
              \draw (\x*1.1, 0) rectangle (\x*1.1+0.8, -\h*0.3);
              \node at (\x*1.1+0.4, -\h*0.15) {\small\h};
          \else
              \draw[dashed, gray] (\x*1.1, 0) rectangle (\x*1.1+0.8, -0.3);
              \node[gray] at (\x*1.1+0.4, -0.15) {\small 0};
          \fi
          \node[above, font=\small] at (\x*1.1+0.4, 0.1) {$\label$};
      }

      \draw[thick] (-5, 0) -- (5.5, 0);

      \node[below, font=\tiny, gray] at (-1*1.1+0.4, -0.55) {\$30.00};
      \node[above, font=\tiny, gray] at (1*1.1+0.4, 0.55) {\$30.01};

      \draw[<->] (-0.3, -2.5) -- (1.1, -2.5);
      \node[below, font=\small] at (0.4, -2.5) {n=1};

      \node[blue!70!black, font=\small\bfseries] at (-3, -3.2) {Bid};
      \node[red!70!black, font=\small\bfseries] at (3.5, -3.2) {Ask};

      \draw[thick, blue!70!black] (-1.1, 0) rectangle (-0.3, 1.5);
      \draw[thick, red!70!black] (1.1, 0) rectangle (1.9, -2.1);
  \end{tikzpicture}
  \caption{Order book with queues indexed relative to the current best
  bid and ask. The best queues $q_{-1}, q_1$ are never empty by
  definition. Deeper queues (dashed) may be empty.}
  \label{fig:lob}
  \end{figure}

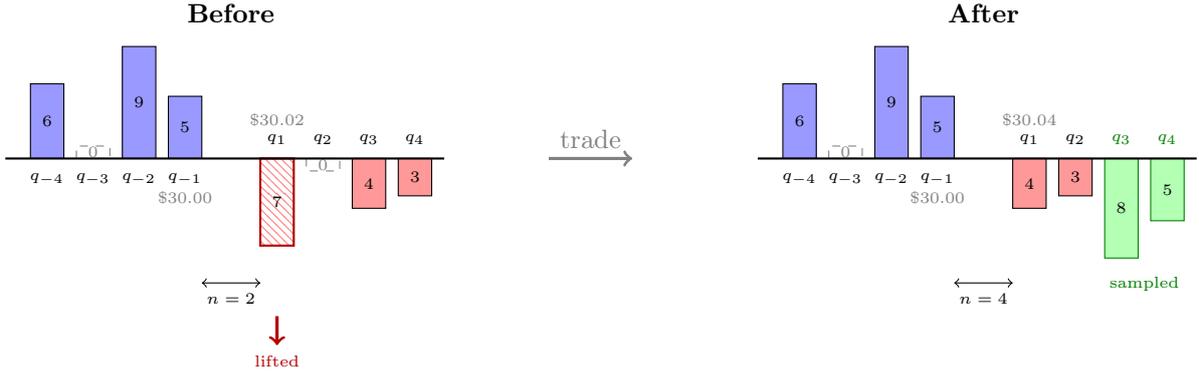
\begin{figure}[H]
  \centering
  \begin{tikzpicture}[scale=0.55]
      \begin{scope}[shift={(-8,0)}]
          \node[font=\small\bfseries] at (0.4, 3.5) {Before};

          \foreach \x/\h/\label in {-4/6/q_{-4}, -3/0/q_{-3}, -2/9/q_{-2}, -1/5/q_{-1}} {
              \ifnum\h>0
                  \fill[blue!40] (\x*1.1, 0) rectangle (\x*1.1+0.8, \h*0.3);
                  \draw (\x*1.1, 0) rectangle (\x*1.1+0.8, \h*0.3);
                  \node at (\x*1.1+0.4, \h*0.15) {\tiny\h};
              \else
                  \draw[dashed, gray] (\x*1.1, 0) rectangle (\x*1.1+0.8, 0.3);
                  \node[gray] at (\x*1.1+0.4, 0.15) {\tiny 0};
              \fi
              \node[below, font=\tiny] at (\x*1.1+0.4, -0.1) {$\label$};
          }
          \node[below, font=\tiny, gray] at (-1*1.1+0.4, -0.55) {\$30.00};

          \fill[red!20, pattern=north west lines, pattern color=red!50]
              (1*1.1, 0) rectangle (1*1.1+0.8, -2.1);
          \draw[red!70!black, thick] (1*1.1, 0) rectangle (1*1.1+0.8, -2.1);
          \node at (1*1.1+0.4, -1.05) {\tiny 7};
          \node[above, font=\tiny] at (1*1.1+0.4, 0.1) {$q_1$};

          \foreach \x/\h/\label in {2/0/q_2, 3/4/q_3, 4/3/q_4} {
              \ifnum\h>0
                  \fill[red!40] (\x*1.1, 0) rectangle (\x*1.1+0.8, -\h*0.3);
                  \draw (\x*1.1, 0) rectangle (\x*1.1+0.8, -\h*0.3);
                  \node at (\x*1.1+0.4, -\h*0.15) {\tiny\h};
              \else
                  \draw[dashed, gray] (\x*1.1, 0) rectangle (\x*1.1+0.8, -0.3);
                  \node[gray] at (\x*1.1+0.4, -0.15) {\tiny 0};
              \fi
              \node[above, font=\tiny] at (\x*1.1+0.4, 0.1) {$\label$};
          }
          \node[above, font=\tiny, gray] at (1*1.1+0.4, 0.55) {\$30.02};

          \draw[thick] (-5, 0) -- (5.5, 0);

          \draw[<->] (-0.3, -3) -- (1.1, -3);
          \node[below, font=\tiny] at (0.4, -3) {$n=2$};

          \draw[->, red!70!black, very thick] (1*1.1+0.4, -3.8) -- (1*1.1+0.4, -4.5);
          \node[below, font=\tiny, red!70!black] at (1*1.1+0.4, -4.5) {lifted};
      \end{scope}

      \draw[->, very thick, gray] (0, 0) -- (2, 0);
      \node[above, font=\small, gray] at (1, 0) {trade};

      \begin{scope}[shift={(10,0)}]
          \node[font=\small\bfseries] at (0.4, 3.5) {After};

          \foreach \x/\h/\label in {-4/6/q_{-4}, -3/0/q_{-3}, -2/9/q_{-2}, -1/5/q_{-1}} {
              \ifnum\h>0
                  \fill[blue!40] (\x*1.1, 0) rectangle (\x*1.1+0.8, \h*0.3);
                  \draw (\x*1.1, 0) rectangle (\x*1.1+0.8, \h*0.3);
                  \node at (\x*1.1+0.4, \h*0.15) {\tiny\h};
              \else
                  \draw[dashed, gray] (\x*1.1, 0) rectangle (\x*1.1+0.8, 0.3);
                  \node[gray] at (\x*1.1+0.4, 0.15) {\tiny 0};
              \fi
              \node[below, font=\tiny] at (\x*1.1+0.4, -0.1) {$\label$};
          }
          \node[below, font=\tiny, gray] at (-1*1.1+0.4, -0.55) {\$30.00};

          \fill[red!40] (1*1.1, 0) rectangle (1*1.1+0.8, -1.2);
          \draw (1*1.1, 0) rectangle (1*1.1+0.8, -1.2);
          \node at (1*1.1+0.4, -0.6) {\tiny 4};
          \node[above, font=\tiny] at (1*1.1+0.4, 0.1) {$q_1$};

          \fill[red!40] (2*1.1, 0) rectangle (2*1.1+0.8, -0.9);
          \draw (2*1.1, 0) rectangle (2*1.1+0.8, -0.9);
          \node at (2*1.1+0.4, -0.45) {\tiny 3};
          \node[above, font=\tiny] at (2*1.1+0.4, 0.1) {$q_2$};

          \fill[green!30] (3*1.1, 0) rectangle (3*1.1+0.8, -2.4);
          \draw[green!50!black] (3*1.1, 0) rectangle (3*1.1+0.8, -2.4);
          \node at (3*1.1+0.4, -1.2) {\tiny 8};
          \node[above, font=\tiny, green!50!black] at (3*1.1+0.4, 0.1) {$q_3$};

          \fill[green!30] (4*1.1, 0) rectangle (4*1.1+0.8, -1.5);
          \draw[green!50!black] (4*1.1, 0) rectangle (4*1.1+0.8, -1.5);
          \node at (4*1.1+0.4, -0.75) {\tiny 5};
          \node[above, font=\tiny, green!50!black] at (4*1.1+0.4, 0.1) {$q_4$};

          \node[above, font=\tiny, gray] at (1*1.1+0.4, 0.55) {\$30.04};

          \draw[thick] (-5, 0) -- (5.5, 0);

          \draw[<->] (-0.3, -3) -- (1.1, -3);
          \node[below, font=\tiny] at (0.4, -3) {$n=4$};

          \node[below, font=\tiny, green!50!black] at (3.5*1.1+0.4, -2.6) {sampled};
      \end{scope}
  \end{tikzpicture}
  \caption{Price move after the best ask $q_1$ is fully depleted by a
  trade. Since $q_2$ was empty, the old $q_3$ becomes the new $q_1$:
  the best ask jumps from \$30.02 to \$30.04 and the mid-price moves
  from \$30.01 to \$30.02, a one-tick upward move. The newly revealed
  deeper queues (green) are sampled from the empirical stationary
  distribution.}
  \label{fig:price_move}
  \end{figure}

  During simulation, we track queues up to $q_{\pm 4}$. When a trade
  depletes the best ask (or bid), all queues shift: the old $q_2$
  becomes the new $q_1$, and so on. The newly revealed deepest queue
  is sampled from the empirical stationary distribution at that level.
  Figure~\ref{fig:price_move} illustrates this reindexing.

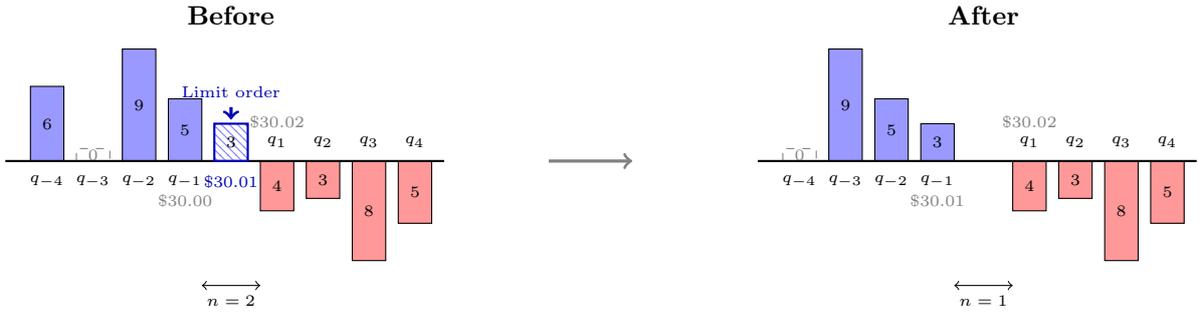
\begin{figure}[H]
  \centering
  \begin{tikzpicture}[scale=0.55]
      \begin{scope}[shift={(-8,0)}]
          \node[font=\small\bfseries] at (0.4, 3.5) {Before};

          \foreach \x/\h/\label in {-4/6/q_{-4}, -3/0/q_{-3}, -2/9/q_{-2}, -1/5/q_{-1}} {
              \ifnum\h>0
                  \fill[blue!40] (\x*1.1, 0) rectangle (\x*1.1+0.8, \h*0.3);
                  \draw (\x*1.1, 0) rectangle (\x*1.1+0.8, \h*0.3);
                  \node at (\x*1.1+0.4, \h*0.15) {\tiny\h};
              \else
                  \draw[dashed, gray] (\x*1.1, 0) rectangle (\x*1.1+0.8, 0.3);
                  \node[gray] at (\x*1.1+0.4, 0.15) {\tiny 0};
              \fi
              \node[below, font=\tiny] at (\x*1.1+0.4, -0.1) {$\label$};
          }
          \node[below, font=\tiny, gray] at (-1*1.1+0.4, -0.55) {\$30.00};

          \foreach \x/\h/\label in {1/4/q_1, 2/3/q_2, 3/8/q_3, 4/5/q_4} {
              \fill[red!40] (\x*1.1, 0) rectangle (\x*1.1+0.8, -\h*0.3);
              \draw (\x*1.1, 0) rectangle (\x*1.1+0.8, -\h*0.3);
              \node at (\x*1.1+0.4, -\h*0.15) {\tiny\h};
              \node[above, font=\tiny] at (\x*1.1+0.4, 0.1) {$\label$};
          }
          \node[above, font=\tiny, gray] at (1*1.1+0.4, 0.55) {\$30.02};

          \draw[thick] (-5, 0) -- (5.5, 0);

          \draw[<->] (-0.3, -3) -- (1.1, -3);
          \node[below, font=\tiny] at (0.4, -3) {$n=2$};

          \fill[blue!20, pattern=north west lines, pattern color=blue!50]
              (0, 0) rectangle (0.8, 0.9);
          \draw[blue!70!black, thick] (0, 0) rectangle (0.8, 0.9);
          \node at (0.4, 0.45) {\tiny 3};
          \node[below, font=\tiny, blue!70!black] at (0.4, -0.1) {\$30.01};
          \draw[->, blue!70!black, very thick] (0.4, 1.3) -- (0.4, 1.0);
          \node[above, font=\tiny, blue!70!black] at (0.4, 1.3) {Limit order};
      \end{scope}

      \draw[->, very thick, gray] (0, 0) -- (2, 0);

      \begin{scope}[shift={(10,0)}]
          \node[font=\small\bfseries] at (0.4, 3.5) {After};

          \fill[blue!40] (-1*1.1, 0) rectangle (-1*1.1+0.8, 0.9);
          \draw (-1*1.1, 0) rectangle (-1*1.1+0.8, 0.9);
          \node at (-1*1.1+0.4, 0.45) {\tiny 3};
          \node[below, font=\tiny] at (-1*1.1+0.4, -0.1) {$q_{-1}$};
          \node[below, font=\tiny, gray] at (-1*1.1+0.4, -0.55) {\$30.01};

          \fill[blue!40] (-2*1.1, 0) rectangle (-2*1.1+0.8, 1.5);
          \draw (-2*1.1, 0) rectangle (-2*1.1+0.8, 1.5);
          \node at (-2*1.1+0.4, 0.75) {\tiny 5};
          \node[below, font=\tiny] at (-2*1.1+0.4, -0.1) {$q_{-2}$};

          \fill[blue!40] (-3*1.1, 0) rectangle (-3*1.1+0.8, 2.7);
          \draw (-3*1.1, 0) rectangle (-3*1.1+0.8, 2.7);
          \node at (-3*1.1+0.4, 1.35) {\tiny 9};
          \node[below, font=\tiny] at (-3*1.1+0.4, -0.1) {$q_{-3}$};

          \draw[dashed, gray] (-4*1.1, 0) rectangle (-4*1.1+0.8, 0.3);
          \node[gray] at (-4*1.1+0.4, 0.15) {\tiny 0};
          \node[below, font=\tiny] at (-4*1.1+0.4, -0.1) {$q_{-4}$};

          \foreach \x/\h/\label in {1/4/q_1, 2/3/q_2, 3/8/q_3, 4/5/q_4} {
              \fill[red!40] (\x*1.1, 0) rectangle (\x*1.1+0.8, -\h*0.3);
              \draw (\x*1.1, 0) rectangle (\x*1.1+0.8, -\h*0.3);
              \node at (\x*1.1+0.4, -\h*0.15) {\tiny\h};
              \node[above, font=\tiny] at (\x*1.1+0.4, 0.1) {$\label$};
          }
          \node[above, font=\tiny, gray] at (1*1.1+0.4, 0.55) {\$30.02};

          \draw[thick] (-5, 0) -- (5.5, 0);

          \draw[<->] (-0.3, -3) -- (1.1, -3);
          \node[below, font=\tiny] at (0.4, -3) {$n=1$};
      \end{scope}
  \end{tikzpicture}
  \caption{A limit order placed inside the spread narrows it. A bid
  limit order at \$30.01 creates a new best bid, shifting all bid
  queues: the spread narrows from $n=2$ to $n=1$.}
  \label{fig:create_event}
  \end{figure}

  Figure~\ref{fig:create_event} shows the reverse situation: a limit
  order placed inside the spread narrows it from the bid side.

\subsection{State projection}
\label{sec:imbalance_proj}

  We project the book onto two quantities: the volume
  imbalance at the best level
  \begin{equation*}
      \text{Imb} = \frac{q_{-1} - q_1}{q_{-1} + q_1} \in [-1, 1]
  \end{equation*}
  and the bid-ask spread $n$ in ticks. The full state
  reduces to $\Phi(\text{LOB}) = (\text{Imb}, n)$.

  The imbalance is discretised into 21 bins of width $0.1$. Negative
  bins are left-closed and labelled by their left edge: an observed
  imbalance of $-0.47$ falls into the $-0.5$ bin via $[-0.5, -0.4)$.
  Positive bins are right-closed and labelled by their right edge:
  $0.13$ falls into the $0.2$ bin via $(0.1, 0.2]$. The central bin at
  $\text{Imb} = 0$ is a point bin capturing exact balance only. Values
  in $(-0.1, 0)$ and $(0, 0.1)$ fall into the $-0.1$ and $0.1$ bins
  respectively. This choice is deliberate as not to mix in the same bin opposite imbalance signs.
  \begin{figure}[ht]
  \centering
  \begin{tikzpicture}[xscale=0.45, yscale=1.1]
      \draw[thick] (-10.5, 0) -- (10.5, 0);

      \foreach \x in {-10,-9,...,10} {
          \draw (\x, -0.15) -- (\x, 0.15);
          \fill (\x, 0) circle (1.5pt);
      }

      \foreach \x/\l in {-10/-1.0, -8/-0.8, -6/-0.6, -4/-0.4, -2/-0.2, 0/0, 2/0.2, 4/0.4, 6/0.6, 8/0.8, 10/1.0} {
          \node[below, font=\scriptsize] at (\x, -0.2) {$\l$};
      }

      \fill[red!20] (-0.3, -0.15) rectangle (0.3, 0.15);
      \fill[red!70!black] (0, 0) circle (2.5pt);

      \fill[blue!70!black] (-4.7, 0) circle (2.5pt);
      \node[above, font=\tiny, blue!70!black] at (-4.7, 0.3) {$-0.47$};
      \draw[->, blue!70!black, thick] (-4.7, -0.3) -- (-5, -0.7);
      \node[below, font=\tiny, blue!70!black] at (-5, -0.8) {bin $-0.5$};

      \fill[green!50!black] (1.3, 0) circle (2.5pt);
      \node[above, font=\tiny, green!50!black] at (1.3, 0.3) {$0.13$};
      \draw[->, green!50!black, thick] (1.3, -0.3) -- (2, -0.7);
      \node[below, font=\tiny, green!50!black] at (2, -0.8) {bin $0.2$};

      \fill[orange!80!black] (-0.3, 0) circle (2.5pt);
      \node[above, font=\tiny, orange!80!black] at (-0.3, 0.3) {$-0.03$};
      \draw[->, orange!80!black, thick] (-0.3, -0.3) -- (-1, -0.7);
      \node[below, font=\tiny, orange!80!black] at (-1, -0.8) {bin $-0.1$};
  \end{tikzpicture}
  \caption{Imbalance binning scheme. Bins have width $0.1$; the $0$ bin
  (highlighted) captures exact balance only. Three example values and
  their bin assignments are shown.}
  \label{fig:imb_bins}
  \end{figure}
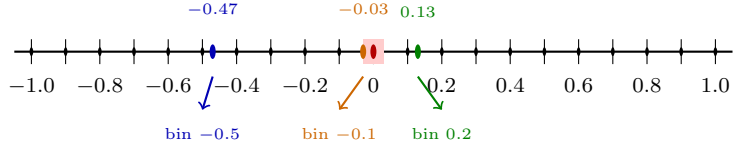

  The bin width choice balances granularity against sufficient sample size.
  One might worry that the point bin at $\text{Imb} = 0$ is too
  restrictive: exact equality $q_{-1} = q_1$ seems unlikely for raw queue
  sizes. However, queue sizes are first normalised by the median event
  size at each level and rounded up to integers, which significantly
  concentrates the distribution. In practice, the $0$ bin is relatively well populated.
  \begin{figure}[ht]
  \begin{tcolorbox}[
    enhanced,
    colback={rgb,255:red,235;green,245;blue,251},
    colframe={rgb,255:red,0;green,102;blue,204},
    arc=6pt, boxrule=0.8pt,
    left=4pt, right=4pt, top=4pt, bottom=4pt
  ]
  \centering
  \includegraphics[width=0.32\textwidth]{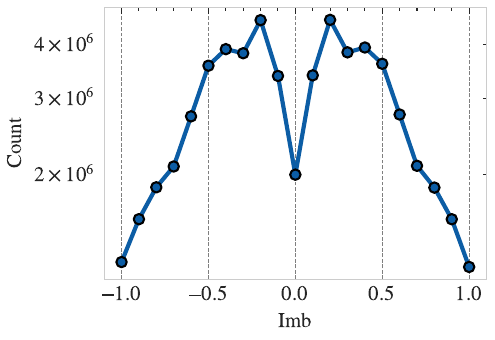}\hfill
  \includegraphics[width=0.32\textwidth]{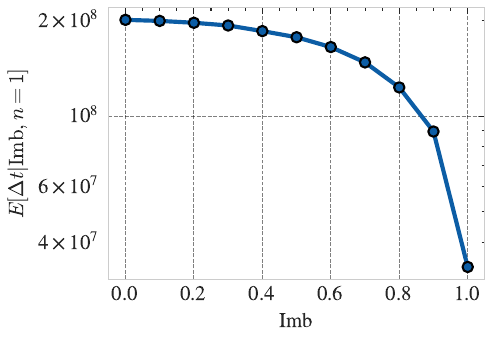}\hfill
  \includegraphics[width=0.32\textwidth]{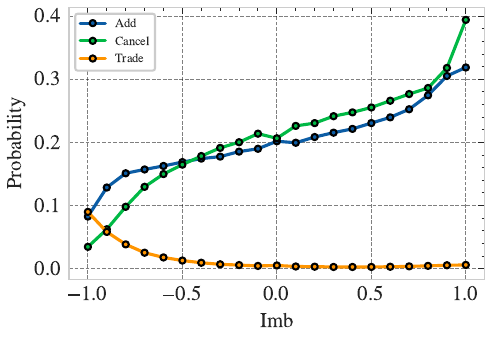}
  \tcblower
  \caption{\textbf{Left:} Observation count per imbalance bin; the $0$ bin is
  more populated than highly imbalanced books.
  \textbf{Centre:} Average inter-event time
  $\mathbb{E}[\Delta t \mid \text{Imb}, n=1]$ per imbalance bin; extreme
  imbalances correspond to faster activity.
  \textbf{Right:} Event probabilities at the best bid $q_{-1}$ given
  imbalance ($n=1$). Positive imbalance means $q_{-1} > q_1$; probabilities
  do not sum to 1 since other queues also receive events. All statistics were calibrated on data spanning $12/2023 \rightarrow 12/2025$.}
  \label{fig:state_projection}
  \end{tcolorbox}
  \end{figure}

  Note that imbalance is a ratio: a book with $q_{-1} = 1, q_1 = 1$ and one
  with $q_{-1} = 50, q_1 = 50$ both map to $\text{Imb} = 0$, despite very
  different dynamics, a single cancellation depletes a queue in the first
  case but is negligible in the second. To address this, one can enrich
  the state with the total resting volume:
  \begin{equation*}
      \Phi(\text{LOB}) = (\text{Imb}, n, \ell) \quad \text{where} \quad
      \ell = q_{-1} + q_1
  \end{equation*}
  In practice, $\ell$ is discretised into a small number of bins (e.g.\ 5
  levels: very low, low, mid, high, very high) defined by empirical
  quantiles of $\ell$ observed in the data. However, we have found
  empirically (Appendix~\ref{app:total_volume}) that this additional
  dimension does not significantly improve the model's reproducing power
  while diluting the observations across a larger number of bins, each
  cell accumulates fewer samples, leading to noisier estimates. We
  therefore use $\Phi(\text{LOB}) = (\text{Imb}, n)$ throughout.

    \subsection{Modelling events}
    \label{sec:events}
\begin{wrapfigure}{r}{0.3\columnwidth}
  \centering
  \vspace{-10pt}
  \includegraphics[width=0.3\columnwidth]{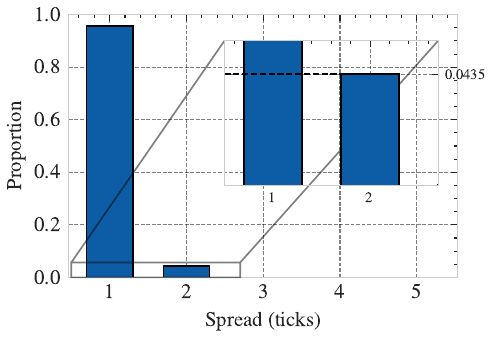}
  \caption{Spread distribution at trade time (PFE).}
  \label{fig:spread_trades}
  \vspace{-10pt}
  \end{wrapfigure} 
  In our model, an event is a tuple $e = (\mathcal{T}, s, i)$ where
  $\mathcal{T}$ is the type (limit order (\textsc{Add}), cancellation
  (\textsc{Cancel}), or market order (\textsc{Trade})),
  $s \in \{-1, +1\}$ is the side, and $i \geq 1$ is the queue level,
  so the target queue is $q_{s \cdot i}$.
  Here, additions and cancellations may target queues up to $q_{\pm 2}$. Trades target only the best queues $q_{\pm 1}$: while trades that
  walk through multiple price levels do occur, they are rare enough
  that we do not model them as a separate event type (special care is
  taken in volume estimation, see Section~\ref{sec:volumes}). In
  practice, most aggressive orders are submitted as marketable limit
  orders priced at the best opposing quote, not as market orders that
  sweep through the book. When the spread $n = 1$, the possible events
  are therefore additions and cancellations at $q_{\pm 1}$ and
  $q_{\pm 2}$, and trades at $q_{\pm 1}$.

  When the spread widens ($n \geq 2$), we introduce two additional event
  types: \textsc{CreateBid} and \textsc{CreateAsk}. A
  \textsc{CreateBid} places a new limit order one tick above the
  current best bid, narrowing the spread from the bid side. Symmetrically,
  a \textsc{CreateAsk} places one tick below the current best ask.
  The model is designed with large-tick assets in mind.
  Figure~\ref{fig:spread_trades} shows the distribution of the spread
  at the time of a trade for PFE: less than 5\% of trades occur when
  the spread is greater than one tick. It is therefore reasonable to
  treat $n \geq 2$ as a transient state resolved exclusively by
  \textsc{Create} events, the spread closes before any other activity
  resumes. For \textsc{Create} events, the queue level is not
  applicable and we set $i = 0$ by convention.

  \subsection{Order volumes}
  \label{sec:volumes}

In the original queue-reactive model, all orders have unit size: each
  event adds or removes exactly one unit from a queue. A ``unit'' is a
  normalisation of the raw volume by a characteristic size. In
  \citet{huang2015simulating}, this is the mean event size; we use the
  median event size (MES) instead, which is more robust to outliers.
  The MES is computed separately for each queue level
  $q_1, q_2, q_3, \ldots$ and symmetrised between bid and ask (so
  $q_{-i}$ and $q_i$ share the same MES). Because the MES differs
  across queue levels, one unit at $q_1$ does not
  represent the same number of shares as one unit at $q_2$. Care
  must be taken when the price moves and queues shift levels: a queue
  that was at level $q_2$ and becomes $q_1$ after a price move must
  have its volume re-expressed in the new level's MES. In practice, order
  sizes vary widely and carry information about the aggressiveness of
  market participants. The role of order sizes has been explored in
  \citep{bodor2024novel}. Moreover, an important mechanism in LOB
  dynamics is when a trade consumes the entire best queue, causing the
  spread to widen. With unit sizes, this can only happen one unit at a
  time, requiring a dedicated event type to capture queue depletion in
  a single step. We introduce random volumes: given the current state
  $\Phi(\text{LOB})$ and event $e$, the order size is drawn from a
  state-dependent distribution:
  \begin{equation*}
    v \sim p(v \mid \Phi(\text{LOB}), e)
  \end{equation*}
  where $p(v \mid \Phi(\text{LOB}), e)$ is estimated as the empirical
  size distribution in each $(\Phi(\text{LOB}), e, q_{\pm i})$ cell,
  symmetrised between bid and ask. Sizes are capped at 50 MES, beyond
  which the empirical density is negligible. Queue depletion then
  arises naturally whenever the best queue is small and a trade of
  moderate size hits it, without any special treatment.

  \paragraph{Volume preprocessing.} Raw event sizes require some care
  before estimation.

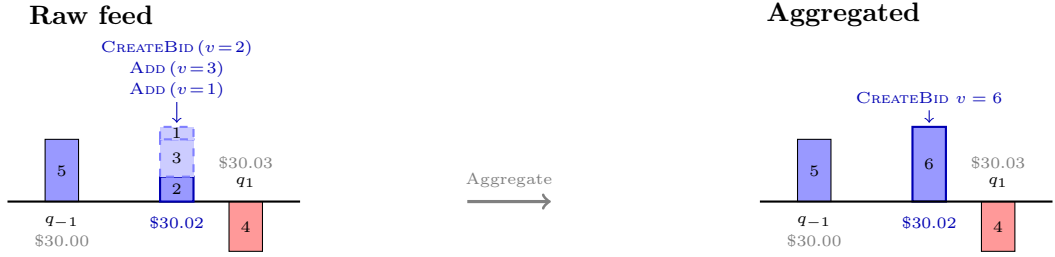
\begin{figure}[H]
  \centering
  \begin{tikzpicture}[scale=0.55]
      \begin{scope}[shift={(-8,0)}]
          \node[font=\small\bfseries] at (0, 4.5) {Raw feed};

          \fill[blue!40] (-1*1.1, 0) rectangle (-1*1.1+0.8, 1.5);
          \draw (-1*1.1, 0) rectangle (-1*1.1+0.8, 1.5);
          \node at (-1*1.1+0.4, 0.75) {\tiny 5};
          \node[below, font=\tiny] at (-1*1.1+0.4, -0.1) {$q_{-1}$};
          \node[below, font=\tiny, gray] at (-1*1.1+0.4, -0.55) {\$30.00};

          \fill[red!40] (3*1.1, 0) rectangle (3*1.1+0.8, -1.2);
          \draw (3*1.1, 0) rectangle (3*1.1+0.8, -1.2);
          \node at (3*1.1+0.4, -0.6) {\tiny 4};
          \node[above, font=\tiny] at (3*1.1+0.4, 0.1) {$q_1$};
          \node[above, font=\tiny, gray] at (3*1.1+0.4, 0.55) {\$30.03};

          \draw[thick] (-2, 0) -- (5, 0);

          \fill[blue!40] (1.5*1.1, 0) rectangle (1.5*1.1+0.8, 0.6);
          \draw[blue!70!black, thick] (1.5*1.1, 0) rectangle (1.5*1.1+0.8, 0.6);
          \node at (1.5*1.1+0.4, 0.3) {\tiny 2};

          \fill[blue!20] (1.5*1.1, 0.6) rectangle (1.5*1.1+0.8, 1.5);
          \draw[blue!50, thick, dashed] (1.5*1.1, 0.6) rectangle (1.5*1.1+0.8, 1.5);
          \node at (1.5*1.1+0.4, 1.05) {\tiny 3};

          \fill[blue!20] (1.5*1.1, 1.5) rectangle (1.5*1.1+0.8, 1.8);
          \draw[blue!50, thick, dashed] (1.5*1.1, 1.5) rectangle (1.5*1.1+0.8, 1.8);
          \node at (1.5*1.1+0.4, 1.65) {\tiny 1};

          \node[font=\tiny, blue!70!black, align=center] at (1.5*1.1+0.4, 3.2)
              {\textsc{CreateBid}\,($v\!=\!2$)\\[1pt]\textsc{Add}\,($v\!=\!3$)\\[1pt]\textsc{Add}\,($v\!=\!1$)};
          \draw[->, blue!70!black] (1.5*1.1+0.4, 2.4) -- (1.5*1.1+0.4, 1.9);

          \node[below, font=\tiny, blue!70!black] at (1.5*1.1+0.4, -0.1) {\$30.02};
      \end{scope}

      \draw[->, very thick, gray] (1, 0) -- (3, 0);
      \node[above, font=\tiny, gray] at (2, 0.1) {Aggregate};

      \begin{scope}[shift={(10,0)}]
          \node[font=\small\bfseries] at (0, 4.5) {Aggregated};

          \fill[blue!40] (-1*1.1, 0) rectangle (-1*1.1+0.8, 1.5);
          \draw (-1*1.1, 0) rectangle (-1*1.1+0.8, 1.5);
          \node at (-1*1.1+0.4, 0.75) {\tiny 5};
          \node[below, font=\tiny] at (-1*1.1+0.4, -0.1) {$q_{-1}$};
          \node[below, font=\tiny, gray] at (-1*1.1+0.4, -0.55) {\$30.00};

          \fill[blue!40] (1.5*1.1, 0) rectangle (1.5*1.1+0.8, 1.8);
          \draw[blue!70!black, thick] (1.5*1.1, 0) rectangle (1.5*1.1+0.8, 1.8);
          \node at (1.5*1.1+0.4, 0.9) {\tiny 6};
          \node[below, font=\tiny, blue!70!black] at (1.5*1.1+0.4, -0.1) {\$30.02};

          \node[font=\tiny, blue!70!black] at (1.5*1.1+0.4, 2.5)
              {\textsc{CreateBid} $v=6$};
          \draw[->, blue!70!black] (1.5*1.1+0.4, 2.2) -- (1.5*1.1+0.4, 1.9);

          \fill[red!40] (3*1.1, 0) rectangle (3*1.1+0.8, -1.2);
          \draw (3*1.1, 0) rectangle (3*1.1+0.8, -1.2);
          \node at (3*1.1+0.4, -0.6) {\tiny 4};
          \node[above, font=\tiny] at (3*1.1+0.4, 0.1) {$q_1$};
          \node[above, font=\tiny, gray] at (3*1.1+0.4, 0.55) {\$30.03};

          \draw[thick] (-2, 0) -- (5, 0);
      \end{scope}
  \end{tikzpicture}
  \caption{Volume aggregation for \textsc{Create} events. The raw feed
  shows a \textsc{CreateBid} at \$30.02 ($v=2$) followed by two
  \textsc{Add} messages at the same price ($v=3$ and $v=1$). We
  aggregate them into a single \textsc{Create} with $v=6$.}
  \label{fig:create_aggregation}
  \end{figure}

  \textsc{Create} events deserve special attention. In the data, a
  level creation is typically followed by a burst of \textsc{Add}
  messages at the same price as other participants join the newly
  created level. If we estimate the \textsc{Create} volume from the
  first message alone, we underestimate the true size of the new
  queue. We therefore aggregate consecutive \textsc{Add} messages at
  the same price into the \textsc{Create} event, stopping as soon as
  any other event is observed (since subsequent additions can no longer
  be reliably attributed to the original creation). The
  individual additions are discarded as separate events and their
  volumes are absorbed into a single \textsc{Create} with the combined
  size.
  Figure~\ref{fig:create_aggregation} illustrates this.

\begin{figure}[H]
  \centering
  \begin{tikzpicture}[scale=0.55]
      \begin{scope}[shift={(-8,0)}]
          \node[font=\small\bfseries] at (0, 4.5) {Raw feed};

          \fill[blue!40] (-1*1.1, 0) rectangle (-1*1.1+0.8, 1.5);
          \draw (-1*1.1, 0) rectangle (-1*1.1+0.8, 1.5);
          \node at (-1*1.1+0.4, 0.75) {\tiny 5};
          \node[below, font=\tiny] at (-1*1.1+0.4, -0.1) {$q_{-1}$};
          \node[below, font=\tiny, gray] at (-1*1.1+0.4, -0.55) {\$30.00};

          \fill[red!40] (1*1.1, 0) rectangle (1*1.1+0.8, -0.6);
          \draw[red!70!black, thick] (1*1.1, 0) rectangle (1*1.1+0.8, -0.6);
          \node at (1*1.1+0.4, -0.3) {\tiny 3};
          
          \fill[red!25] (1*1.1, -0.6) rectangle (1*1.1+0.8, -1.2);
          \draw[red!50, thick, dashed] (1*1.1, -0.6) rectangle (1*1.1+0.8, -1.2);
          \node at (1*1.1+0.4, -0.9) {\tiny 2};

          \fill[red!25] (1*1.1, -1.2) rectangle (1*1.1+0.8, -1.5);
          \draw[red!50, thick, dashed] (1*1.1, -1.2) rectangle (1*1.1+0.8, -1.5);
          \node at (1*1.1+0.4, -1.35) {\tiny 1};

          \node[above, font=\tiny] at (1*1.1+0.4, 0.1) {$q_1$};
          \node[above, font=\tiny, gray] at (1*1.1+0.4, 0.55) {\$30.01};

          \node[font=\tiny, red!70!black, align=center] at (1*1.1+0.4, -2.8)
              {\textsc{Trade}\,($v\!=\!3$)\\[1pt]\textsc{Trade}\,($v\!=\!2$)\\[1pt]\textsc{Trade}\,($v\!=\!1$)};
          \draw[->, red!70!black] (1*1.1+0.4, -2.1) -- (1*1.1+0.4, -1.6);

          \draw[thick] (-2, 0) -- (4, 0);
      \end{scope}

      \draw[->, very thick, gray] (1, 0) -- (3, 0);
      \node[above, font=\tiny, gray] at (2, 0.1) {Aggregate};

      \begin{scope}[shift={(10,0)}]
          \node[font=\small\bfseries] at (0, 4.5) {Aggregated};

          \fill[blue!40] (-1*1.1, 0) rectangle (-1*1.1+0.8, 1.5);
          \draw (-1*1.1, 0) rectangle (-1*1.1+0.8, 1.5);
          \node at (-1*1.1+0.4, 0.75) {\tiny 5};
          \node[below, font=\tiny] at (-1*1.1+0.4, -0.1) {$q_{-1}$};
          \node[below, font=\tiny, gray] at (-1*1.1+0.4, -0.55) {\$30.00};

          \fill[red!40] (1*1.1, 0) rectangle (1*1.1+0.8, -1.5);
          \draw[red!70!black, thick] (1*1.1, 0) rectangle (1*1.1+0.8, -1.5);
          \node at (1*1.1+0.4, -0.75) {\tiny 6};
          \node[above, font=\tiny] at (1*1.1+0.4, 0.1) {$q_1$};
          \node[above, font=\tiny, gray] at (1*1.1+0.4, 0.55) {\$30.01};

          \node[font=\tiny, red!70!black] at (1*1.1+0.4, -2.3)
              {\textsc{Trade} $v=6$};
          \draw[->, red!70!black] (1*1.1+0.4, -2.0) -- (1*1.1+0.4, -1.6);

          \draw[thick] (-2, 0) -- (4, 0);
      \end{scope}
  \end{tikzpicture}
  \caption{Volume aggregation for \textsc{Trade} events. The raw feed
  shows three \textsc{Trade} messages at \$30.01 ($v=3$, $v=2$, $v=1$)
  corresponding to a single aggressive order hitting three resting
  orders. We aggregate them into a single \textsc{Trade} with $v=6$.}
  \label{fig:trade_aggregation}
  \end{figure}
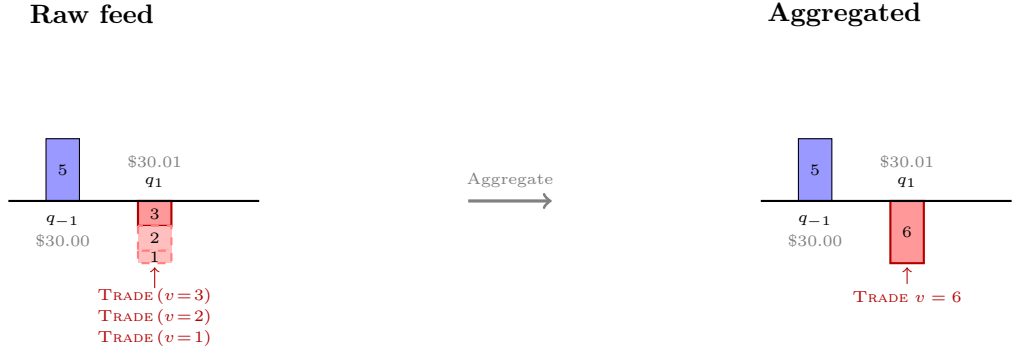

  Trade messages require similar care. When an aggressive order
  executes against multiple resting orders at the best price, the feed
  reports each fill as a separate \textsc{Trade} message sharing the
  same timestamp. We aggregate all such consecutive messages into a
  single trade with their combined volume
  (Figure~\ref{fig:trade_aggregation}). More rarely, a large
  aggressive order may walk through several price levels; we aggregate
  these as well and record only the total traded volume. Walk-the-book
  events are infrequent enough that modelling them as a distinct event
  type is not worthwhile: in our framework every trade is treated as a
  marketable limit order, and any residual volume that is not filled
  reappears as a limit order on the opposite side.

  \subsection{Sampling}

  Although events only target queues up to $q_{\pm 2}$, we track the
  book up to $q_{\pm 4}$. The deeper queues $q_{\pm 3}$ and
  $q_{\pm 4}$ are never modified directly, but they become relevant
  after a price move: when a trade depletes the best queue, all queues
  shift and a formerly deep queue may become the new best. Tracking
  these levels ensures that the book state after a price move is
  informed by data rather than resampled entirely.

  Let $\mathcal{E}(\Phi(\text{LOB}))$ denote the set of possible
  events for a given book state as described in
  Section~\ref{sec:events}. As per \citet{huang2015simulating} the model is a Markov jump
  process: in state $\Phi(\text{LOB})$, each possible event
  $e \in \mathcal{E}(\Phi(\text{LOB}))$ has its own exponential clock
  with rate $\lambda^e(\Phi(\text{LOB}))$. By defining the total rate and event probabilities:
  \begin{equation*}
      \Lambda(\Phi(\text{LOB})) = \sum_{e \in \mathcal{E}(\Phi(\text{LOB}))} \lambda^e(\Phi(\text{LOB})), \qquad
      p^e(\Phi(\text{LOB})) = \frac{\lambda^e(\Phi(\text{LOB}))}{\Lambda(\Phi(\text{LOB}))}
  \end{equation*}
  Sampling the next event reduces to two independent draws:
  \begin{equation*}
      \Delta t \sim \text{Exp}\!\big(\Lambda(\Phi(\text{LOB}))\big), \quad
      e^* \sim \sum_{e \in \mathcal{E}(\Phi(\text{LOB}))} p^e(\Phi(\text{LOB})) \, \delta_e
  \end{equation*}

  The simulation then proceeds as follows.

  \begin{algorithm}[H]
  \DontPrintSemicolon
  \SetAlgoLined
  \KwIn{Estimated parameters $\hat{p}^e$, $\hat{\Lambda}$, $\hat{p}(v \mid \cdot)$; initial order book; $t \leftarrow 0$}
  \While{simulating}{
    Compute current state $\Phi(\text{LOB})$\;
    Sample event $e^* \sim \sum_{e \in \mathcal{E}(\Phi(\text{LOB}))} p^e(\Phi(\text{LOB}))\,\delta_e$\;
    Sample waiting time $\Delta t \sim \mathrm{Exp}\!\big(\Lambda(\Phi(\text{LOB}))\big)$; set $t \leftarrow t + \Delta t$\;
    Sample volume $v^* \sim p(v \mid \Phi(\text{LOB}),\, e^*)$\;
    Apply event $e^*$ with size $v^*$ to the order book\;
  }
  \caption{Queue-Reactive simulation loop}
  \end{algorithm}

  \subsection{Parameter estimation}
  \label{sec:estimation}

  Our dataset consists of transitions
  $\mathcal K = \{(\Delta t_k,\, e_k,\, \Phi(\text{LOB}_k))\}_{k=1}^{N}$, where $N$ is the number of observed events. Define :
  \begin{equation*}
      \mathcal{K}(\Phi(\text{LOB})) = \{k : \Phi(\text{LOB}_k) = \Phi(\text{LOB})\}, \qquad
      \mathcal{K}(\Phi(\text{LOB}), e) = \{k \in \mathcal{K}(\Phi(\text{LOB})) : e_k = e\}
  \end{equation*}
  All parameters are estimated by maximum likelihood. The event
  probabilities are the empirical frequencies in each
  $\Phi(\text{LOB})$ bin:
  \begin{equation*}
      \hat{p}^e(\Phi(\text{LOB})) =
      \frac{\#\mathcal{K}(\Phi(\text{LOB}), e)}
           {\#\mathcal{K}(\Phi(\text{LOB}))}
  \end{equation*}

  The total intensity is the inverse of the mean waiting time:
  \begin{equation*}
      \hat{\Lambda}(\Phi(\text{LOB})) = \left(
      \frac{1}{\#\mathcal{K}(\Phi(\text{LOB}))}
      \sum_{k \in \mathcal{K}(\Phi(\text{LOB}))} \Delta t_k
      \right)^{-1}
  \end{equation*}

  After estimation, all statistics are symmetrised between bid and ask:
  the probability of a bid event at imbalance $+x$ is averaged with
  that of the corresponding ask event at imbalance $-x$, and similarly
  for waiting times and size parameters. Any residual asymmetry in the
  estimated parameters would translate into a systematic price drift on
  a macroscopic scale, which is not a feature we want the baseline
  model to exhibit. Symmetrisation eliminates this artefact, reduces
  the effective number of parameters by half, and produces smoother
  estimates.

  We verify in Appendix~\ref{app:stability} that parameters estimated
  on disjoint six-month windows remain consistent across the sample
  period. Event probabilities are remarkably stable; inter-event times
  preserve their functional form but shift in level, consistent with
  time-varying market activity that the practitioner captures through
  periodic re-estimation.

  All estimation is performed on
  \href{https://databento.com/docs/schemas-and-data-formats/mbp-10?historical=python&live=python&reference=python}{Databento MBP-10}
  data covering December 2023 to December 2025. To avoid the peculiarities of the open and close we discard the 
  first and last $30$ minutes of each trading day, effectively maintaining the 10:00--15:30 window.

\subsection{Baseline validation}
\label{sec:validation}

We evaluate whether the simulator reproduces the marginal statistics
that a practitioner would check before trusting its output: the mix of
event types, the book states under which trades occur, overall activity
levels, price volatility, and the return distribution. These statistics
test progressively deeper aspects of the model, from per-event fidelity
to emergent price dynamics.

We calibrate the QR model on PFE using two years $12/2023\rightarrow12/2025$ of data.
We let the model run for an equivalent of $1000$ trading hours and compare
the resulting simulation to our data.

\begin{figure}[ht]
\begin{tcolorbox}[
  enhanced,
  colback={rgb,255:red,235;green,245;blue,251},
  colframe={rgb,255:red,0;green,102;blue,204},
  arc=6pt,
  boxrule=0.8pt,
  left=4pt, right=4pt, top=4pt, bottom=4pt
]
\centering
\begin{subfigure}[t]{0.32\textwidth}
  \centering
  \includegraphics[width=\textwidth]{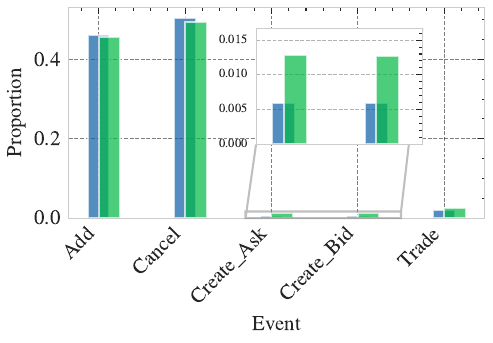}
  \caption{Event-type distribution.}
  \label{fig:event_distrib}
\end{subfigure}\hfill
\begin{subfigure}[t]{0.32\textwidth}
  \centering
  \includegraphics[width=\textwidth]{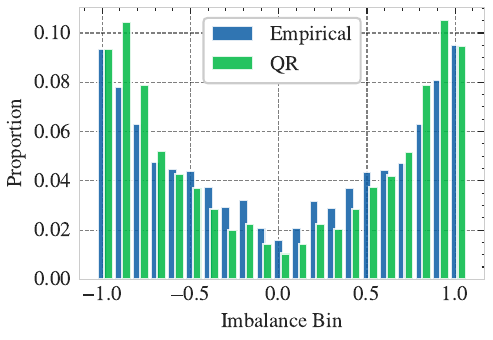}
  \caption{Imbalance before trades.}
  \label{fig:imb_before_trade}
\end{subfigure}\hfill
\begin{subfigure}[t]{0.32\textwidth}
  \centering
  \includegraphics[width=\textwidth]{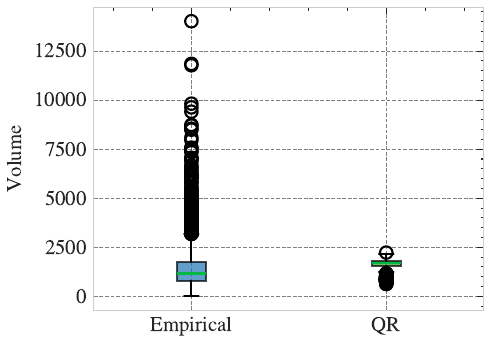}
  \caption{Hourly traded volume.}
  \label{fig:hourly_volume}
\end{subfigure}

\vspace{6pt}

\begin{subfigure}[t]{0.32\textwidth}
  \centering
  \includegraphics[width=\textwidth]{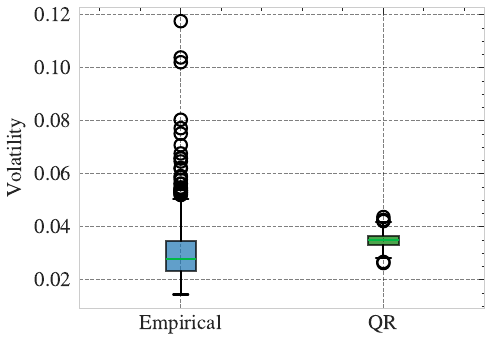}
  \caption{5-min realized volatility.}
  \label{fig:volatility}
\end{subfigure}\hfill
\begin{subfigure}[t]{0.32\textwidth}
  \centering
  \includegraphics[width=\textwidth]{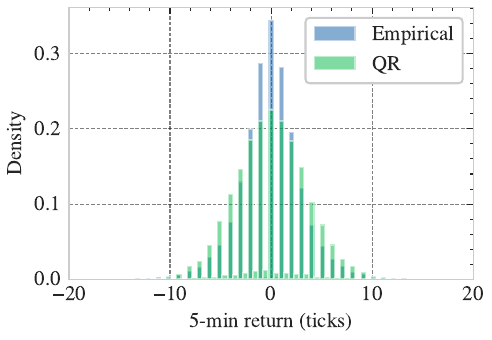}
  \caption{5-min returns distribution.}
  \label{fig:returns_5m}
\end{subfigure}\hfill
\begin{subfigure}[t]{0.32\textwidth}
  \centering
  \includegraphics[width=\textwidth]{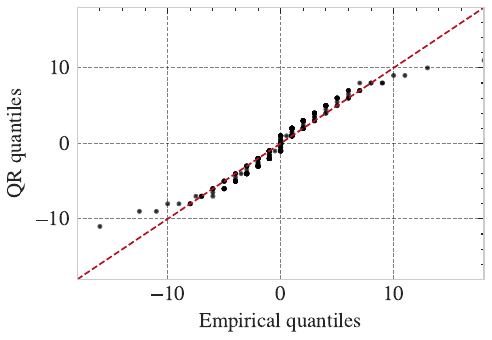}
  \caption{5-min returns QQ plot.}
  \label{fig:returns_5m_qq}
\end{subfigure}
\tcblower
\caption{Baseline validation statistics. Empirical data (blue) vs.\ QR simulation (green/orange) for PFE.}
\label{fig:validation_grid}
\end{tcolorbox}
\end{figure}

  \paragraph{Event-type distribution.}
  Figure~\ref{fig:event_distrib} compares the relative frequencies of the different possible events between the empirical data and the
  QR simulation.  The model reproduces the dominant Add/Cancel balance
  faithfully: both account for roughly 97\% of all events, while trades
  represent approximately 2\% and level-creation events less than 1\%.

  \paragraph{Imbalance before trades.}
  Figure~\ref{fig:imb_before_trade} shows the distribution of the signed
  queue imbalance observed immediately before
  each trade.  Both distributions exhibit a U-shape, confirming
  that trades are most likely when one side of the book is heavily depleted.
  The QR model slightly overweights the extreme imbalance bins relative to
  the empirical data, consistent with the absence of any feedback mechanism
  between trades and the order flow.

  \paragraph{Activity levels.}
  As a sanity check, we confirm that the simulated hourly traded volume
  (Figure~\ref{fig:hourly_volume}) is consistent with the fitted
  intensities. The QR model concentrates around the empirical average,
  which is expected by construction. The narrower spread of the
  simulated distribution reflects the absence of day-to-day variation
  in market conditions: the model captures average behavior, not regime
  variation.

  \paragraph{Five-minute realized volatility.}
  The original QR model of \citet{huang2015simulating} reported
  systematic volatility underestimation, which warranted the
  introduction of a $\theta_\text{reinit}$ parameter to model exogenous
  price moves. Under our framework, we find that volatility matches on
  average what we see in the data (Figure~\ref{fig:volatility}). We
  attribute this to two differences: the introduction of random
  volumes~\citep{bodor2024novel}, which allow a single trade to
  deplete an entire queue, and
  our order book representation, which discards the fixed reference
  price $p_{\text{ref}}$ and the associated reindexing mechanism of the
  original model.
  As a measure of volatility we compute the following:
  \begin{equation*}
    \sigma_{\text{day}} = \sqrt{\frac{1}{T - 1} \sum_{t=1}^{T-1} (p_{t+1} - p_t)^2}
\end{equation*}
where $p_t$ is the last traded price in the $t$-th 5-minute bin and $T = 66$ is the number of bins per day (6.5 trading hours minus 30 minutes from open and 30 minutes from close, divided into 5-minute intervals).

\paragraph{Five-minute returns.}
    Figure~\ref{fig:returns_5m} compares the distribution of 5-minute
    mid-to-mid returns.
    The QQ plot (right) confirms that the bulk of the distribution is
    well captured: the points align closely with the diagonal for returns
    within $\pm 5$ ticks. The tails are lighter than in the
    data. This is expected: the QR model is ergodic and the midprice is asymptotically Brownian~\citep{huang2015ergodicity}, so it cannot produce
    the large price moves driven by exogenous shocks or persistent
    order flow.

\subsection*{Limitations}

The baseline model has two main limitations that we address in the
following sections.

\begin{figure}[H]
    \centering
    \includegraphics[width=0.48\columnwidth]{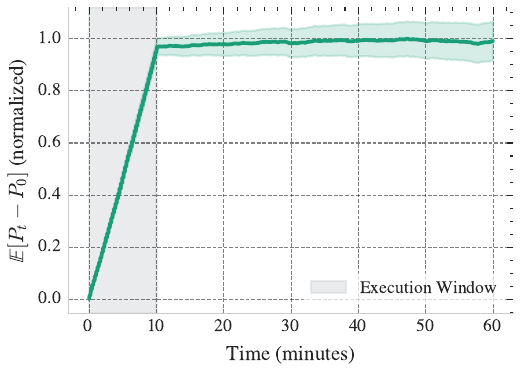}\hfill
    \includegraphics[width=0.48\columnwidth]{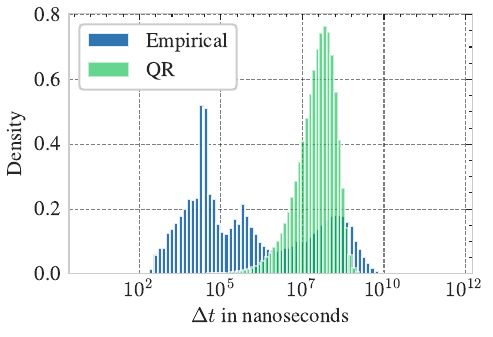}
    \caption{\textbf{Left:} Average price path around a simulated
    metaorder in the baseline QR model, no impact is observed.
    \textbf{Right:} Distribution of inter-event times $\Delta t$
    (blue is empirical, green is QR).}
    \label{fig:no_impact_metaorder}
\end{figure}

\paragraph{Inter-event times.}
As shown in Figure~\ref{fig:no_impact_metaorder}, the empirical
distribution of $\Delta t$ is far from the exponential type assumed by the
QR model, a discrepancy that, to our knowledge, has not been addressed
in the queue-reactive literature. This is not merely a statistical
curiosity: the real distribution reveals that a significant fraction
of events arrive nearly simultaneously, too fast to be sequential
reactions. For any strategy that depends on being first to act on a
signal, the timing of competing orders is what determines whether it
gets filled. An exponential clock, which spreads events uniformly in
time, cannot capture this clustering and systematically
overestimates fill rates.

\paragraph{Market impact.}
The QR model has no mechanism for market impact beyond the effect of queue modifications: the transition
probabilities depend only on the current book state, not on the
history of trades. Passive impact is well reproduce but When we simulate an aggressive metaorder in the baseline model,
the average price path shows no concave rise during execution and no
reversion afterwards (Figure~\ref{fig:no_impact_metaorder}). This is
fundamentally at odds with empirical observations, where metaorders impact prices in a specific, path-dependent manner \citep{bouchaud2008marketsslowlydigestchange,toth2011anomalous,durin2023two,muhle2026unified}.

Let us also mention two other non Markovian features of empirical data, that can easily be incorporated in a QR framework if considered important.

\paragraph{Order flow autocorrelation.}
Empirically, trade signs exhibit long memory: their
autocorrelation decays slowly, as a power law. The QR simulation
reproduces short-lag autocorrelation from the book state alone, but
the decay is much faster. This is not necessarily a feature the user aim to reproduce:
the long memory arises from the strategic behavior of institutional type
participants, not from book microstructure. If this property is of key interest for the user, it can easily be obtained using a Hawkes-type component in the market flow, see for example \citep{bodor2024novel}.

\paragraph{Path-dependent queue survival time.}
The mean time until a small best queue is fully depleted is, in practice, not only state-dependent but also path-dependent. The QR model treats a queue of size 5 identically regardless of how that state was reached. However, in reality, a queue that has been gradually reduced from 50 likely consists of orders submitted by participants with strong conviction. Such orders are plausibly less likely to be cancelled than more recently placed orders. This effect could be captured by enriching the model’s state space so as to account explicitly for path dependence.

\section{Queue-Reactive model with races}
\label{sec:races}

\subsection{Exchange latency structure}
\label{sec:fingerprint}

\begin{figure}[H]
    \centering
    \includegraphics[width=0.48\columnwidth]{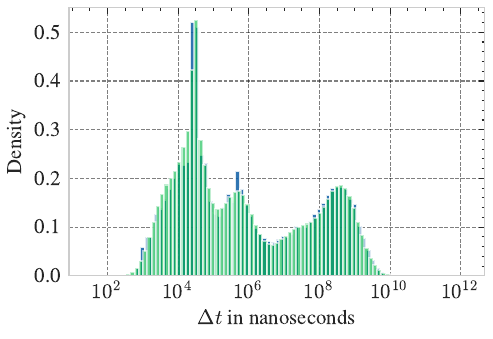}\hfill
    \includegraphics[width=0.48\columnwidth]{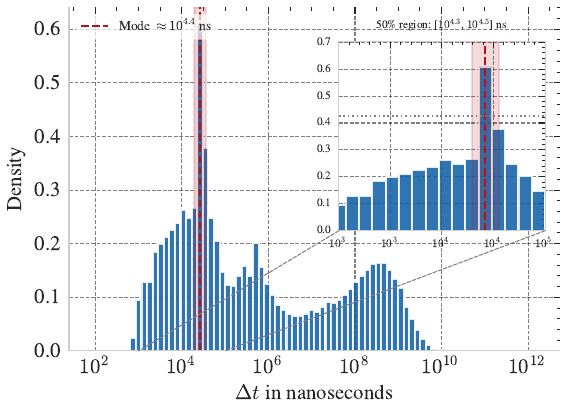}
    \caption{\textbf{Left:} Distribution of inter-event times $\Delta t$
    in $\log_{10}$ of nanoseconds (blue: empirical, green: GMM fit).
    \textbf{Right:} Zoom into the mode, highlighting the region where
    density drops to $70\%$ of its peak value.}
    \label{fig:gmm_fit}
\end{figure}

  When we examine the empirical distribution of inter-event times
  $\Delta t$ in $\log_{10}$ space (Figure~\ref{fig:gmm_fit}), a
  striking feature appears: across all tickers, a consistent mode
  around $\log_{10}(\Delta t) \approx 4.47$, corresponding to
  approximately $29$ microseconds. Zooming into the peak region, the
  interval where density drops to $70\%$ of the maximum is remarkably
  tight: $[4.3, 4.7]$ in $\log_{10}$ space, or roughly $[20, 50]$
  microseconds.

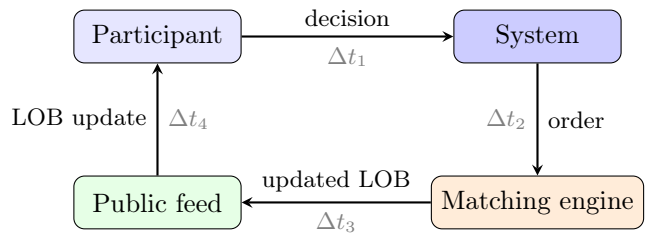
\begin{figure}[H]
  \centering
  \begin{tikzpicture}[
      >=stealth,
      box/.style={draw, fill=gray!20, rounded corners=4pt, minimum width=2.2cm,
                   minimum height=0.7cm, align=center, font=\small},
  ]
      \node[box, fill=blue!10] (PART) at (0, 2.2) {Participant};
      \node[box, fill=blue!20] (SYS)  at (5, 2.2) {System};
      \node[box, fill=orange!15] (ME)   at (5, 0)   {Matching engine};
      \node[box, fill=green!10] (FEED) at (0, 0)   {Public feed};

      \draw[->, thick] (PART.east) -- (SYS.west)
          node[midway, above, font=\footnotesize] {decision}
          node[midway, below, font=\footnotesize, gray] {$\Delta t_1$};

      \draw[->, thick] (SYS.south) -- (ME.north)
          node[midway, right, font=\footnotesize] {order}
          node[midway, left, font=\footnotesize, gray] {$\Delta t_2$};

      \draw[->, thick] (ME.west) -- (FEED.east)
          node[midway, above, font=\footnotesize] {updated LOB}
          node[midway, below, font=\footnotesize, gray] {$\Delta t_3$};

      \draw[->, thick] (FEED.north) -- (PART.south)
          node[midway, left, font=\footnotesize] {LOB update}
          node[midway, right, font=\footnotesize, gray] {$\Delta t_4$};

  \end{tikzpicture}
  
  \caption{The exchange round-trip loop. The participant receives a
  book update from the feed ($\Delta t_4$), makes a decision
  ($\Delta t_1$), and sends an order to the matching engine
  ($\Delta t_2$). The matching engine processes and publishes the
  update ($\Delta t_3$). The round-trip latency is
  $\delta = \Delta t_2 + \Delta t_3 + \Delta t_4$; the measured
  inter-event time is $\Delta t = \Delta t_1 + \delta$.}
  \label{fig:roundtrip}
  \end{figure}

  The timestamps we observe are exchange timestamps: the time at
  which the matching engine receives each order. The
  exchange has many internal timestamps, but the one exposed on the
  public feed can safely be assumed to be consistent across messages.
  Figure~\ref{fig:roundtrip} illustrates the full cycle: a book
  update leaves the matching engine, travels to the participant, the
  participant processes it and sends an order back, and the matching
  engine records the result.

  The full cycle decomposes into four legs
  (Figure~\ref{fig:roundtrip}): $\Delta t_1$ is the participant's
  decision time, $\Delta t_2$ is the network latency from the
  participant's system to the matching engine, $\Delta t_3$ covers the
  matching engine's processing and publication to the feed, and
  $\Delta t_4$ is the network latency from the feed back to the
  participant. We define the \emph{round-trip latency}
  $\delta = \Delta t_2 + \Delta t_3 + \Delta t_4$, which is the time
  between a participant sending an order and seeing the resulting
  update on the feed. The measured inter-event time is then
  $\Delta t = \Delta t_1 + \delta$. For collocated participants the
  network legs are effectively identical across firms, so $\delta$ is
  shared. Since high-frequency systems have decision times
  $\Delta t_1$ of a few microseconds, the mode at $\approx 29\,\mu$s in
  the empirical $\Delta t$ distribution is naturally explained: it
  corresponds to $\delta$ with a negligible $\Delta t_1$. We model
  $\delta$ as a random time concentrated around this mode.

  Roughly $25\%$ of events occur at $\Delta t < \delta$, too
  fast to be a reaction to the preceding event on the public
  feed. These events can in particular be explained as correlated responses
  to a shared signal: when multiple participants observe the same
  directional cue, such as the order book imbalance, they rush to
  trade. The sharp peak at $\Delta t \approx \delta$ corresponds to the
  first order of a race, the fastest possible reaction. Subsequent
  survivors arrive shortly after, separated only by the matching
  engine's processing latency, which produces the cluster of fast
  events just below the peak. However, some events at
  $\Delta t < \delta$ are not race-related at all: participants also
  submit orders asynchronously for reasons unrelated to the most recent
  book update, and these orders may arrive close together by chance.
  Moreover, the public feed only reveals race winners, so the true
  extent of competitive activity is larger than what the data shows.

\begin{wrapfigure}{r}{0.45\columnwidth}
      \centering
      \vspace{-10pt}
      \includegraphics[width=0.45\columnwidth]{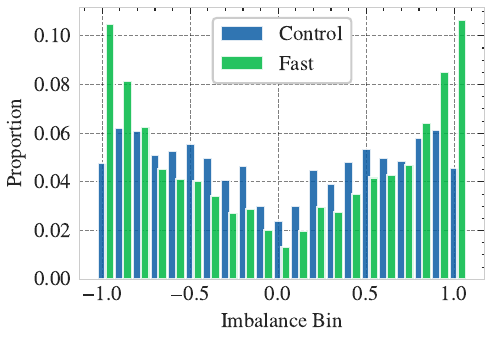}
      \caption{Distribution of imbalance before trades.
      Unconditional (Control $\Delta t > \delta$) vs.\ fast events
      ($\Delta t \approx \delta$).}
      \label{fig:imb_before_trade_fast}
      \vspace{-10pt}
  \end{wrapfigure}
  Figure~\ref{fig:imb_before_trade_fast} compares the distribution of
  imbalance immediately before trades, split by inter-event time.
  Trades that arrive as an immediate reaction ($\Delta t \approx
  \delta$), i.e.\ the first event in a burst, are
  significantly more concentrated at extreme imbalances than trades
  arriving after a longer pause ($\Delta t > \delta$). This is
  consistent with the race phenomenon studied
  in~\citet{aquilina2022quantifying}: when one side of the book is
  heavily depleted, multiple participants converge on the same
  directional view and race to act. The $25\%$ figure, however, only
  partially reflects true races. On the one hand, we only observe race
  winners on the public feed, so the actual number of competing
  participants is higher. On the other hand, many of these fast events
  are simply noise, orders that happen to be processed close together
  without any causal link, since participants also submit orders
  asynchronously for reasons unrelated to the most recent book update.

\subsection{Fitting inter-event times}
\label{sec:fitting_times}

To incorporate realistic timing into the model, we can
replace the exponential clock with the empirical distribution by
exploiting Bayes rule:
\begin{equation*}
  p(\Delta t, e \mid \Phi(\text{LOB})) \propto p(e \mid \Phi(\text{LOB})) \; p(\Delta t \mid \Phi(\text{LOB}), e)
\end{equation*}
The first factor is the event category already estimated in
Section~\ref{sec:estimation}. For the timing component
$p(\Delta t \mid \Phi(\text{LOB}), e)$, one can use the empirical
distribution directly or fit a parametric model. We detail a Gaussian
mixture approach in Appendix~\ref{app:gmm} that reproduces the
empirical distribution of the  $\Delta t$  faithfully
(Figure~\ref{fig:gmm_fit}, left).

Importantly, replacing the exponential clock with the empirical
timing distribution does not alter the simulator's event trajectory:
the book state, event sequence, and direction of price moves are
determined entirely by the event probabilities and the impact
mechanism. The clock only matters when a strategy needs to know
whether its order arrived before or after a competing one.

\subsection{Fill modeling in latency races}
\label{sec:fill_model}

  The latency structure described above has a direct consequence for
  anyone using this simulator to evaluate a high-frequency strategy.
  Consider for example a strategy that monitors the order book imbalance and submits
  a market order when the signal exceeds a threshold. If the
  inter-event time following its order is $\Delta t > \delta$, the
  strategy's order arrived in isolation and can reasonably expect a fill.
  But when $\Delta t < \delta$, the situation is different: the next
  event was not a reaction to the strategy's order, it was a
  simultaneous response to the same signal. In this regime, the strategy
  is in a \emph{latency race} with other participants who observed the
  same cue and acted at the same time.

  The practical implication is that fill probabilities should not always supposed to be one. When the signal is strong enough to trigger the
  strategy, it is likely strong enough to trigger competitors as well.
  Fill probability should decrease with signal strength: at high frequency, a stronger
  signal is probably shared with more racers, reducing the chance that any single
  participant is first in the queue. 

  Modeling the race outcome precisely: who wins, with which
  probability, and how that probability depends on the state of the
  book, is extremely challenging with public feed data alone. The
  public feed neither reveal the identities of competing
  participants, nor the orders that did not make it through the matching engine. Fitting a
  structural model of race dynamics would require proprietary exchange
  data that is not generally available.

  We therefore settle for a simple proxy: if the next QR event
  has $\Delta t > \delta$, the strategy fills. Otherwise, the fill happens with some probability and can even be partial. So a 
  fill probability $p_{fill}(\Delta t)$ activates when $\Delta t < \delta$. This probability may depend on $\Delta t$ but also on the magnitude of the signal triggering the race and can be calibrated by
the user from own proprietary data
  on missed orders.

\section{Impact-aware Queue Reactive}
\label{sec:impact}

\subsection{Market impact mechanism}
\label{sec:impact_mechanism}

When an investor executes a large order, known as \emph{metaorder}, it is typically
  split into many child orders spread over time. The accumulated order flow pushes the price on average in the direction of the trade: buys push it up, sells push it down. Empirically, the impact during execution follows a concave rise, well approximated by a
  square-root law, see the references above. After execution ends,
  the price partially reverts: only a fraction of the peak impact
  persists as permanent impact, the rest is temporary and decays over
  minutes to hours.

  The QR model has no memory of past order flow: event probabilities
  depend only on the current book state. A metaorder moves the price
  during execution, but once it stops the model has no mechanism to
  revert the price. We introduce a feedback mechanism to reproduce
  these dynamics.

  \paragraph{Impact state.}
  Let $\phi_t$ be a running state variable that accumulates signed
  trade flow and decays over time:
  \begin{equation}\label{eq:impact_kernel}
    \phi_t = \sum_{k:\, t_k < t} G(t - t_k)\;\varepsilon_k\;\sqrt{V_k}
  \end{equation}
  where $t_k$ is the time of the $k-$th transaction, $\varepsilon_k = +1$ for ask-side (buy) trades,
  $\varepsilon_k = -1$ for bid-side (sell) trades, $V_k$ is the
  trade size, and $G$ is a non-negative decay kernel specified below.

  \paragraph{Single-sided biasing.}
  The state $\phi_t$ biases trade probabilities toward mean reversion,
  but only on the side that \emph{opposes} the accumulated flow:
  \begin{equation}\label{eq:bias}
   \left\{\begin{aligned}
    &p^{\text{bid trade}}(\cdot) \propto p_0^{\text{bid trade}}(\cdot)\cdot
      e^{\,m\,[\phi_t]^+} \\
    &p^{\text{ask trade}}(\cdot) \propto p_0^{\text{ask trade}}(\cdot)\cdot
      e^{\,m\,[-\phi_t]^+}
      \end{aligned}\right.
  \end{equation}
  where $p_0^{\text{bid trade}}$ and $p_0^{\text{ask trade}}$ denote the baseline probabilities estimated from data
  Section~\ref{sec:estimation}, $[x]^+ = \max(x,0)$, and $m > 0$ is a
  scaling parameter. When $\phi_t > 0$ (recent net buying), only bid
  (sell) trades are made more likely; when $\phi_t < 0$ (recent net
  selling), only ask (buy) trades are boosted. Note that it is possible to
  use different multipliers $m^+$ and $m^-$ for the bid and ask sides
  if one believes buy and sell orders have
  asymmetric impact, as can be  the case in equity markets.

  \paragraph{Kernel and calibration.}
  The feedback mechanism is flexible: the kernel $G$ and the multiplier
  $m$ can be calibrated to any impact criterion that matters to the
  practitioner. It can be a theoretical benchmark, empirical metaorder data, or
  a maximum-likelihood fit to observed price dynamics (see
  Appendix~\ref{app:mle_impact} for the latter). Here we describe our
  own choice, which targets the theoretical impact profile of
  \citet{jusselin2020noarbitrage}.

\begin{wrapfigure}{r}{0.45\columnwidth}
      \centering
      \vspace{-10pt}
      \includegraphics[width=0.43\columnwidth]{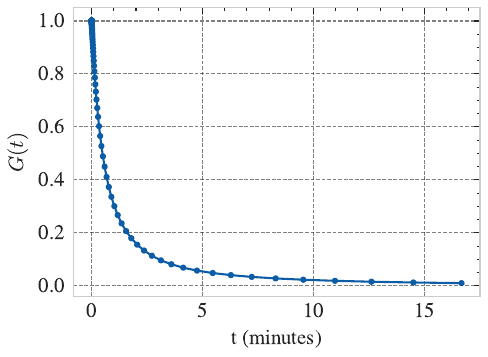}
      \caption{The decay kernel $G(t) = (1 + t/\tau)^{-3/2}$ with
        $\tau = 50$\,s, plotted over a 17-minute window.}
      \label{fig:decay_kernel}
      \vspace{-50pt}
  \end{wrapfigure}

  Since $\phi_t$ acts as a force on the price ($dP/dt \propto -\phi_t$),
  matching an impact that decays as $t^{-1/2}$ after execution requires
  $G(t) \propto t^{-3/2}$, consistent with the no-arbitrage constraints
  of \citet{jusselin2020noarbitrage}. We use
  \begin{equation*}
    G(t) = \left(1 + \frac{t}{\tau}\right)^{-3/2},
  \end{equation*}
  where $\tau$ controls the crossover from flat to power-law decay
  (Figure~\ref{fig:decay_kernel}). The approximation of $G$ as a
  sum of exponentials, which enables $O(1)$ per-event updates, is
  detailed in Appendix~\ref{app:kernel_params}.

  We set $\tau = 50$\,s, which produces a reasonable kernel shape
  (Figure~\ref{fig:decay_kernel}). A more principled approach is to
  estimate $\tau$ by maximum likelihood from observed price dynamics;
  we present this in Appendix~\ref{app:mle_impact} and find
  qualitatively similar results.

  To calibrate $m$, we choose a typical metaorder profile (TWAP,
  10\% of hourly volume, $T = 10$\,min execution, each child order
  of size 2, observed for 60\,min) and a theoretical impact shape we
  wish to reproduce:
  \begin{equation*}
    I_{\text{target}}(t) =
    \begin{cases}
      \sqrt{t/T} & t \leq T \\[4pt]
      \sqrt{t/T} - \sqrt{t/T - 1} & t > T.
    \end{cases}
  \end{equation*}
  We find $m$ by binary search over 100{,}000 Monte Carlo paths,
  minimizing MSE against $I_{\text{target}}$. This yields
  $m = 0.036$. These parameters are used throughout the case studies
  that follow.

\begin{figure}[H]
  \begin{tcolorbox}[
    enhanced,
    colback={rgb,255:red,235;green,245;blue,251},
    colframe={rgb,255:red,0;green,102;blue,204},
    arc=6pt,
    boxrule=0.8pt,
    left=4pt, right=4pt, top=4pt, bottom=4pt
  ]
      \centering
      \includegraphics[width=0.48\columnwidth]{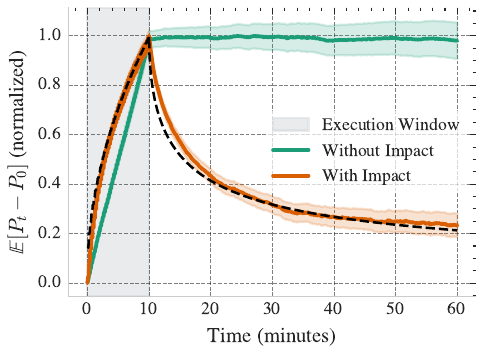}\hfill
      \includegraphics[width=0.48\columnwidth]{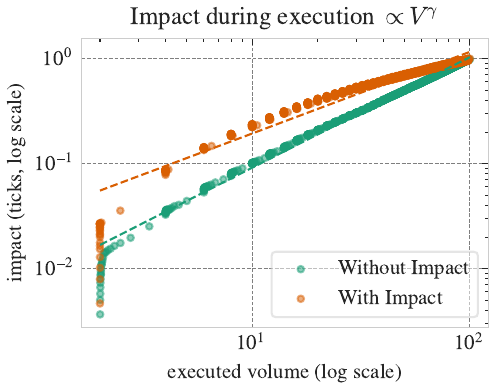}
  \tcblower
      \caption{\textbf{Left:} Average price impact of a metaorder with and without the
        feedback mechanism, normalized by peak value.
        \textbf{Right:} Impact during execution versus executed volume in
        log-log scale.}
      \label{fig:impact}
  \end{tcolorbox}
  \end{figure}

  \paragraph{Results.}
  Figure~\ref{fig:impact} shows the average price path over 500{,}000
  simulated metaorders. Without impact, the price stays at its peak
  after execution. With the feedback mechanism, it reverts and
  stabilizes at a fraction of the peak, consistent with the partial
  reversion observed empirically.

  Figure~\ref{fig:impact} (right) plots the impact during execution
  versus executed volume on a log-log scale. Without feedback, impact
  grows linearly with volume. With the power-law kernel, the growth is
  concave.

  We stress that the choice of target profile is illustrative: the
  kernel structure is flexible enough to match any concave impact
  profile, not just the one predicted by
  \citet{jusselin2020noarbitrage}. In practice, calibrating $\tau$
  and $m$ requires either metaorder data or a theoretical target.
  When proprietary execution data is available, one can calibrate
  directly to the observed impact curve; absent such data, a
  theoretical benchmark like the one used here provides a reasonable
  default.

\section{Case studies}
\label{sec:trading}

With the impact mechanism and timing extensions in place, we
  demonstrate their effect by embedding two strategies in the
  simulator: a mid-frequency driven by a non-order book based alpha
  signal, and a high-frequency strategy that exploits order book
  imbalance and is subject to latency races.

\subsection{Mid-frequency strategy}
\label{sec:mf_case}

  \paragraph{Alpha signal.}
  We model a mid-frequency trading signal as an Ornstein--Uhlenbeck
  process $\alpha_t$ with mean-reversion rate $\kappa$ and diffusion
  coefficient $\sigma$:
  \begin{equation*}
    d\alpha_t = -\kappa\,\alpha_t\,dt + \sigma\,dW_t.
  \end{equation*}
The signal enters the model through a combined bias term
  \begin{equation*}
    b_t = m\,\phi_t - \lambda\,\alpha_t
  \end{equation*}
  where $\lambda > 0$ is a scaling coefficient. The bias $b_t$ enters
  the trade probabilities exactly as in~\eqref{eq:bias}: the impact
  component $m\,\phi_t$ creates mean-reverting pressure, while the
  signal component $-\lambda\,\alpha_t$ tilts probabilities in the
  direction of the forecast. The two terms compete, impact pushes the
  price back while informed flow pushes it forward, and the net bias
  $b_t$ determines the prevailing direction.

  \begin{wrapfigure}{r}{0.45\columnwidth}
      \centering
      \vspace{-10pt}
      \includegraphics[width=0.43\columnwidth]{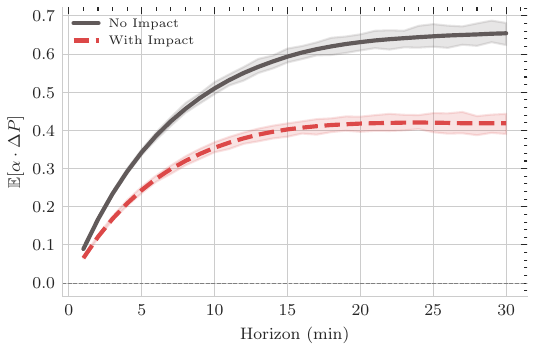}
      \caption{Signal predictiveness: $\mathbb{E}[\alpha \cdot \Delta P(h)]$
        as a function of horizon $h$ (in minutes).}
      \label{fig:alpha_pred}
      \vspace{-20pt}
  \end{wrapfigure}

  Figure~\ref{fig:alpha_pred} confirms that the signal is indeed
  predictive in the simulation: $\mathbb{E}[\alpha \cdot \Delta P(h)]$
  is positive and increasing in the horizon $h$, both with and without
  impact. With impact, predictiveness is reduced at all horizons,
  reflecting the cost of the strategy's own footprint in the market.
  Notably, the two curves diverge after a few minutes and the
  with-impact curve plateaus around $h = 20$\,min, suggesting that
  accumulated self-impact gradually exhausts the signal's edge.
  Confidence bands are 95\% bootstrap confidence intervals.

  \paragraph{Strategy description.}
  The strategy trades whenever $|\alpha_t|$ exceeds a threshold
  $\theta$, buying at the best ask when $\alpha_t > \theta$ and
  selling at the best bid when $\alpha_t < -\theta$. Two risk
  parameters constrain its behavior:
  \begin{itemize}
    \item \textbf{Max inventory} $\bar{I}$: the strategy cannot
      accumulate a position beyond $\bar{I}$ shares in either
      direction.
    \item \textbf{Queue size} $q_{\max}$: the size of each
      individual order, capped so as not to exceed the available
      quantity at the best level.
  \end{itemize}
  As a proxy for execution latency and cooldown between trades, we
  impose that a QR event (a regular market order book event) occurs
  between any two consecutive strategy trades. This prevents the
  strategy from firing multiple orders in rapid succession and
  ensures that the book has time to replenish between fills.

\begin{figure}[H]
  \begin{tcolorbox}[
    enhanced,
    colback={rgb,255:red,235;green,245;blue,251},
    colframe={rgb,255:red,0;green,102;blue,204},
    arc=6pt,
    boxrule=0.8pt,
    left=4pt, right=4pt, top=4pt, bottom=4pt
  ]
      \centering
      \includegraphics[width=0.48\columnwidth]{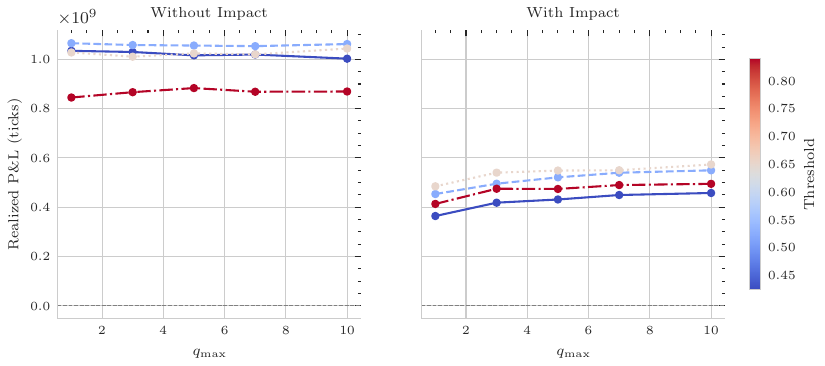}\hfill
      \includegraphics[width=0.48\columnwidth]{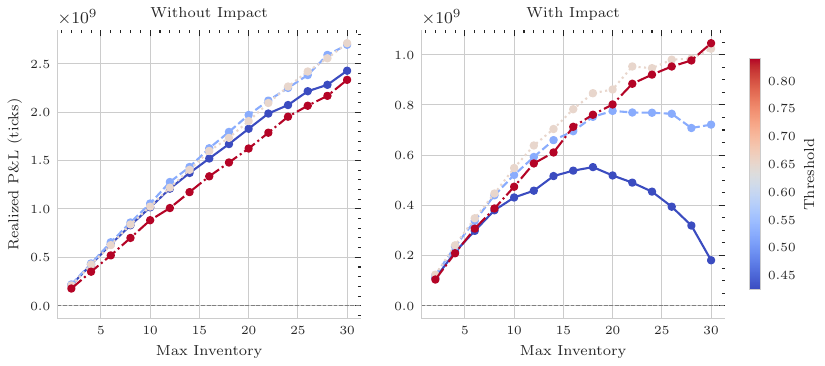}
  \tcblower
      \caption{\textbf{Left:} Realized P\&L versus order size $q_{\max}$;
        the curves flatten quickly, confirming that available depth is
        the binding constraint.
        \textbf{Right:} P\&L versus max inventory, with threshold
        $\theta$ as color gradient. Without impact, more inventory is
        always better; with impact, a trade-off emerges.}
      \label{fig:pnl}
  \end{tcolorbox}
  \end{figure}

  \paragraph{Results.}
  The order size $q_{\max}$ has little effect on P\&L in our setting
  (Figure~\ref{fig:pnl}, left): since it is capped by the available
  depth at the best level, which is typically small, increasing
  $q_{\max}$ beyond this depth does not change behavior. We fix
  $q_{\max} = 5$ throughout and focus on the effect of maximum
  inventory.

  Figure~\ref{fig:pnl} shows realized P\&L as a function of max
  inventory for several values of the threshold $\theta$. Without
  impact, P\&L increases monotonically with max inventory: the more
  the strategy is allowed to accumulate, the more edge it captures.
  With impact, there is a clear trade-off: beyond a certain inventory
  level, the additional trades generate enough market impact to erode
  the signal's edge, and P\&L flattens or declines. The threshold
  $\theta$ controls selectivity, higher thresholds trade less often
  but on stronger signals, shifting the curves upward at the cost of
  lower overall volume.

\subsection{High-frequency strategy}
\label{sec:hft_case}

  \paragraph{Strategy description.}
  The strategy monitors the order book imbalance and submits a market
  order whenever the imbalance exceeds a threshold of $0.85$,
  buying at the ask (selling at the bid) when the book is tilted
  in the corresponding direction. Because the strategy reacts to
  a signal that is visible to all participants, its latency is
  drawn from $\delta$, the exchange round-trip delay described
  in Section~\ref{sec:fingerprint}. Fill is determined here by a simple
 version of the rule of Section~\ref{sec:fill_model}: if the next QR
  event has $\Delta t^{\text{QR}} < \delta$, the strategy's order
  does not execute (it loses the race); otherwise, it fills.

  Two parameters govern risk: a maximum inventory $\bar{I}$ and
  an order size $q_{\max}$. We sweep over both and run the
  strategy under two impact regimes:
  \begin{enumerate}
    \item \textbf{Without self-impact:} the impact mechanism is
      active (other participants' trades move prices via $\phi_t$),
      but the strategy's own fills are excluded, i.e.\ we set $\varepsilon_k=0$ for the strategy's trades in~\eqref{eq:impact_kernel}.
    \item \textbf{With self-impact:} the strategy's trades feed
      back into $\phi_t$ on the same footing as every other trade.
  \end{enumerate}
  The gap between the two regimes isolates the cost of the
  strategy's own footprint in the market.

\begin{figure}[H]
  \begin{tcolorbox}[
    enhanced,
    colback={rgb,255:red,235;green,245;blue,251},
    colframe={rgb,255:red,0;green,102;blue,204},
    arc=6pt,
    boxrule=0.8pt,
    left=4pt, right=4pt, top=4pt, bottom=4pt
  ]
      \centering
      \includegraphics[width=0.48\columnwidth]{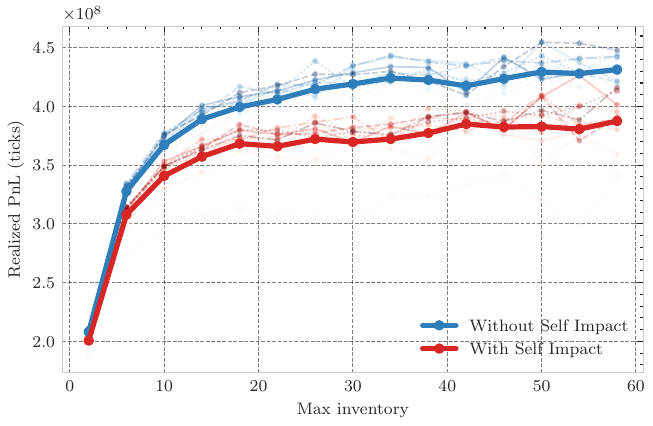}\hfill
      \includegraphics[width=0.48\columnwidth]{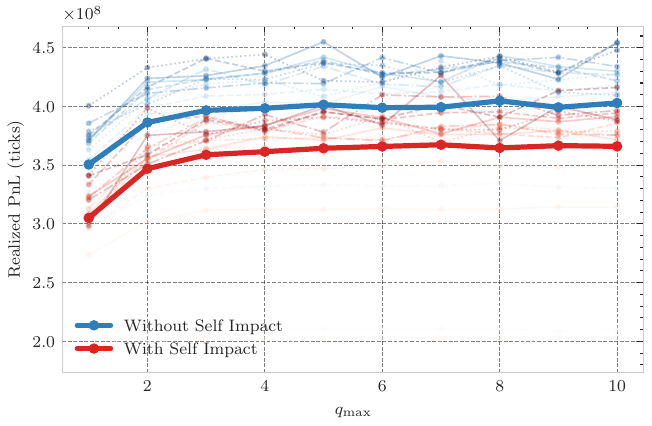}
  \tcblower
      \caption{Realized P\&L of the HFT strategy.
        \textbf{Left:} versus max inventory; faded lines show
        individual $q_{\max}$ values.
        \textbf{Right:} versus $q_{\max}$; faded lines show
        individual max inventory values.
        Bold lines are averages. The gap between the two curves
        measures the cost of the strategy's own impact.}
      \label{fig:hft_pnl}
  \end{tcolorbox}
  \end{figure}

  \paragraph{Results.}
  Figure~\ref{fig:hft_pnl} shows realized P\&L as a function of
  max inventory $\bar{I}$ (left) and order size $q_{\max}$
  (right). Faded lines correspond to individual parameter values;
  bold lines are averages. Both regimes exhibit concave growth in
  $\bar{I}$, but the self-impact curve flattens earlier and at a
  lower level. The vertical gap between the two bold lines is the
  P\&L lost to the strategy's own impact, a cost that grows with
  inventory and that a simulator ignoring self-impact would miss
  entirely. Sliced by $q_{\max}$ (right), both curves flatten
  quickly: the available depth at the best level is the binding
  constraint, so increasing $q_{\max}$ beyond a few units has
  little additional effect. The self-impact cost is roughly
  constant across $q_{\max}$, confirming that it is driven by the
  cumulative position rather than by individual order size.

  These results illustrate why self-impact modeling matters for
  HFT backtesting. A simulator that excludes the strategy's own
  trades from the impact kernel will systematically overstate
  profitability, precisely in the aggressive parameterization
  regime where a practitioner most needs an accurate assessment.

\bibliographystyle{apalike}
\bibliography{references}

\newpage
\appendix

\section{Parameter stability}
\label{app:stability}

\begin{figure}[H]
    \centering
    \includegraphics[width=\textwidth]{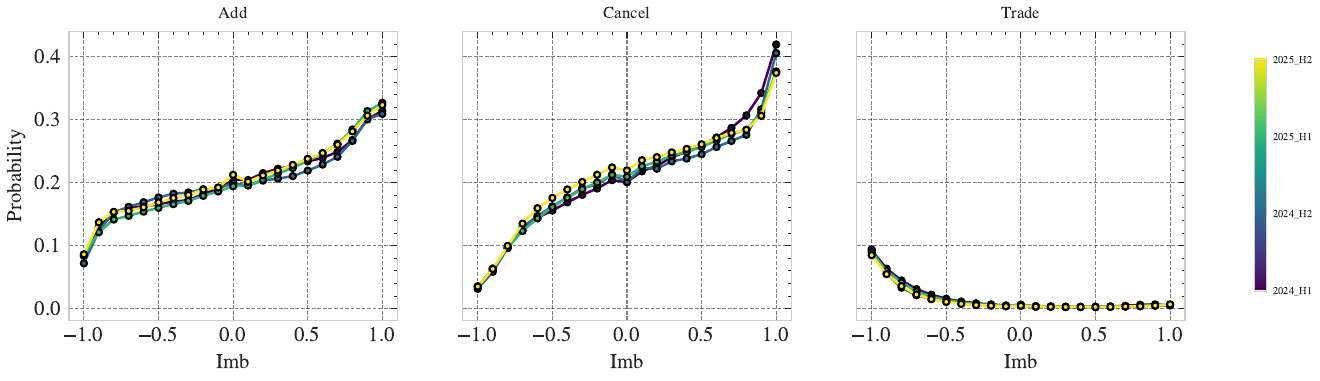}
    \caption{Event probabilities at the best bid ($q_{-1}$) per imbalance
    bin, estimated on disjoint six-month windows (PFE, $n=1$).}
    \label{fig:probs_over_time}
\end{figure}

\begin{wrapfigure}{r}{0.45\columnwidth}
    \centering
    \vspace{-10pt}
    \includegraphics[width=0.43\columnwidth]{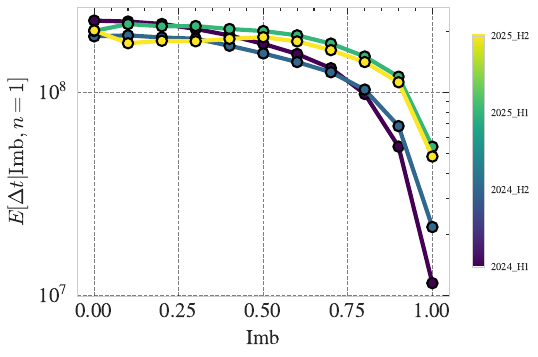}
    \caption{Average inter-event time
    $\mathbb{E}[\Delta t \mid \text{Imb}, n=1]$ per imbalance bin,
    estimated on disjoint six-month windows (PFE).}
    \label{fig:dt_over_time}
    \vspace{-10pt}
\end{wrapfigure}
To assess whether the estimated parameters reflect stable microstructural
features rather than artefacts of a specific sample period, we re-estimate
the model on four disjoint six-month windows (2024~H1, 2024~H2, 2025~H1,
2025~H2) and compare the results.

Figure~\ref{fig:probs_over_time} shows event probabilities at the best
bid as a function of imbalance for each window. The curves are virtually
indistinguishable across all four periods, confirming that the conditional
event structure is a stable property of the market.

Inter-event times (Figure~\ref{fig:dt_over_time}) preserve their
shape, activity increases monotonically with imbalance
magnitude, but the level shifts slightly, with later windows
exhibiting somewhat longer times at extreme imbalances.
This is consistent with time-varying overall market activity rather
than model instability: the \emph{structure} of the parameters is
stable, while their \emph{scale} tracks market conditions.

\section{Effect of total resting volume}
\label{app:total_volume}

We examine whether enriching the state projection with the total resting
volume $\ell = q_{-1} + q_1$ improves the model. We discretise $\ell$ into
five bins defined by empirical quantiles and compare event probabilities,
observation counts, and inter-event times across levels.

\begin{figure}[H]
  \centering
  \includegraphics[width=\textwidth]{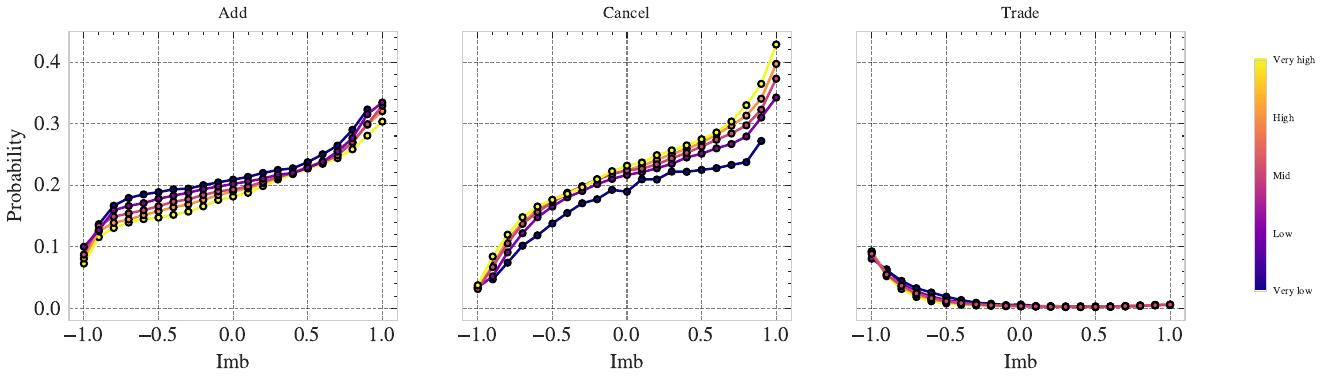}
  \caption{Event probabilities at the best bid ($q_{-1}$) per imbalance
  bin, stratified by total queue level $\ell$ (PFE, $n=1$). The curves
  are relatively similar, suggesting that $\ell$ adds little
  predictive power beyond imbalance alone.}
  \label{fig:probs_total_best}
\end{figure}

Figure~\ref{fig:probs_total_best} shows event probabilities at the best
bid, stratified by $\ell$. The curves are relatively similar across all
levels: the conditional event distribution depends primarily on imbalance,
not on the absolute size of the book.

\begin{figure}[H]
  \centering
  \includegraphics[width=0.48\columnwidth]{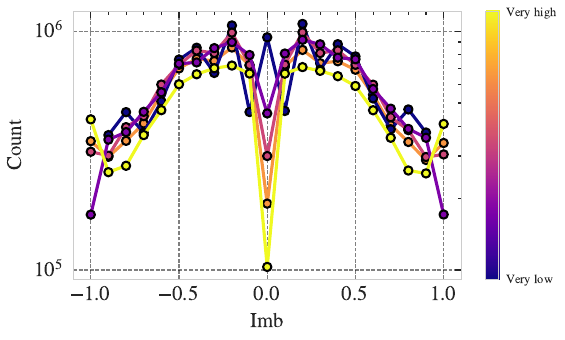}\hfill
  \includegraphics[width=0.48\columnwidth]{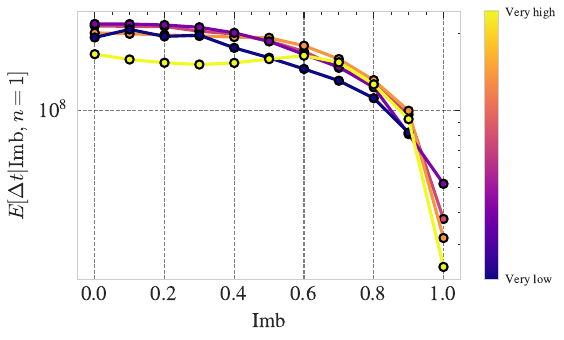}
  \caption{\textbf{Left:} Observation count per imbalance bin, stratified
  by total queue level $\ell$ (PFE, $n=1$).
  \textbf{Right:} Average inter-event time per imbalance bin, stratified
  by $\ell$. The curves largely overlap, confirming that $\ell$ adds
  little beyond imbalance alone.}
  \label{fig:count_imb_total_best}
\end{figure}

Figure~\ref{fig:count_imb_total_best} shows the observation count per
imbalance bin for each level of $\ell$. While the overall shape is
preserved, the counts are diluted across five bins, with the $0$ bin
in particular becoming somewhat sparse for large $\ell$.

\section{Baseline validation statistics for INTC, VZ, and T}
\label{app:validation_tickers}

\begin{figure}[H]
\begin{tcolorbox}[
  enhanced,
  colback={rgb,255:red,235;green,245;blue,251},
  colframe={rgb,255:red,0;green,102;blue,204},
  arc=6pt, boxrule=0.8pt,
  left=4pt, right=4pt, top=4pt, bottom=4pt
]
\centering
\begin{subfigure}[t]{0.32\textwidth}
  \centering
  \includegraphics[width=\textwidth]{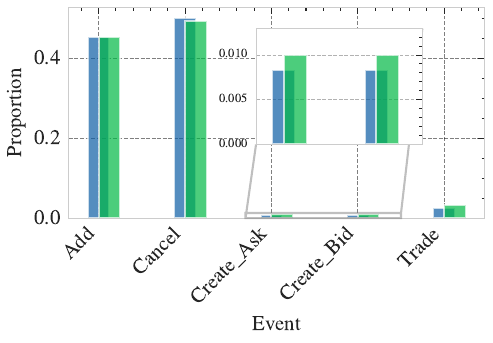}
  \caption{Event-type distribution.}
\end{subfigure}\hfill
\begin{subfigure}[t]{0.32\textwidth}
  \centering
  \includegraphics[width=\textwidth]{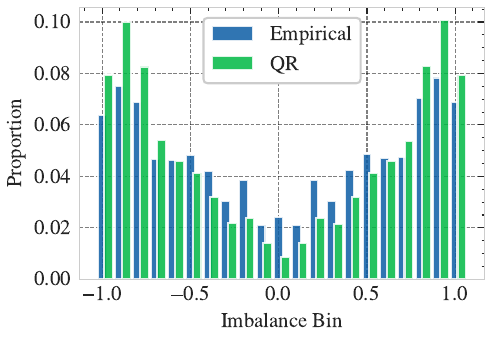}
  \caption{Imbalance before trades.}
\end{subfigure}\hfill
\begin{subfigure}[t]{0.32\textwidth}
  \centering
  \includegraphics[width=\textwidth]{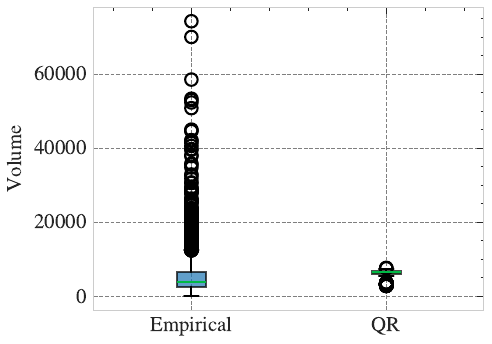}
  \caption{Hourly traded volume.}
\end{subfigure}

\vspace{6pt}

\begin{subfigure}[t]{0.32\textwidth}
  \centering
  \includegraphics[width=\textwidth]{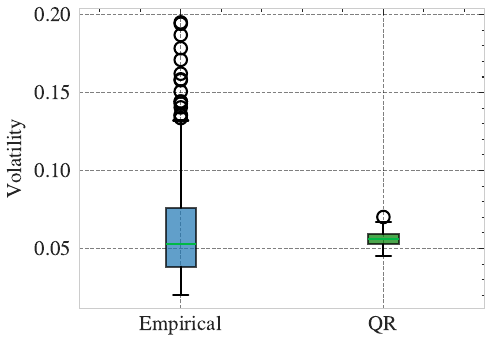}
  \caption{5-min realized volatility.}
\end{subfigure}\hfill
\begin{subfigure}[t]{0.32\textwidth}
  \centering
  \includegraphics[width=\textwidth]{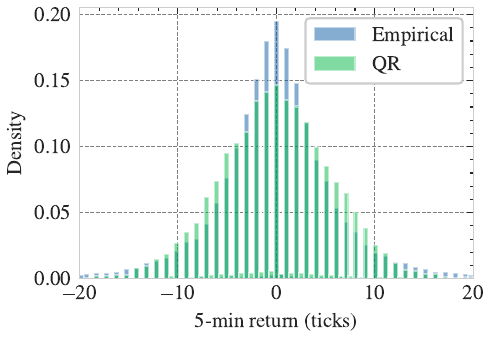}
  \caption{5-min returns distribution.}
\end{subfigure}\hfill
\begin{subfigure}[t]{0.32\textwidth}
  \centering
  \includegraphics[width=\textwidth]{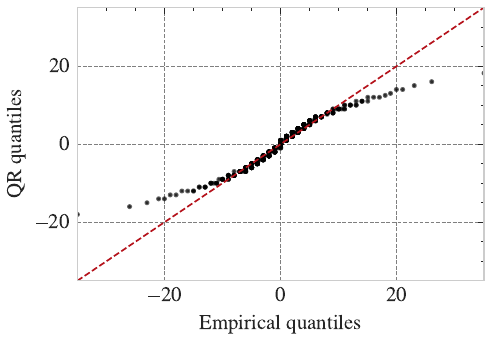}
  \caption{5-min returns QQ plot.}
\end{subfigure}
\tcblower
\caption{Baseline validation statistics. Empirical data (blue) vs.\ QR simulation (green/orange) for INTC.}
\label{fig:validation_INTC}
\end{tcolorbox}
\end{figure}

\begin{figure}[H]
\begin{tcolorbox}[
  enhanced,
  colback={rgb,255:red,235;green,245;blue,251},
  colframe={rgb,255:red,0;green,102;blue,204},
  arc=6pt, boxrule=0.8pt,
  left=4pt, right=4pt, top=4pt, bottom=4pt
]
\centering
\begin{subfigure}[t]{0.32\textwidth}
  \centering
  \includegraphics[width=\textwidth]{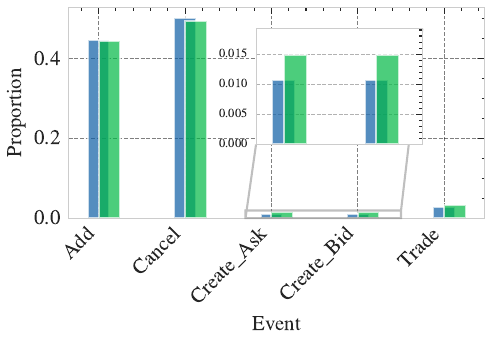}
  \caption{Event-type distribution.}
\end{subfigure}\hfill
\begin{subfigure}[t]{0.32\textwidth}
  \centering
  \includegraphics[width=\textwidth]{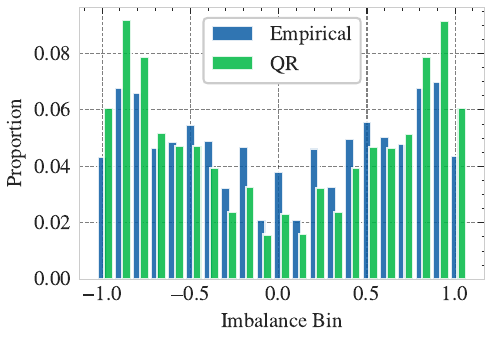}
  \caption{Imbalance before trades.}
\end{subfigure}\hfill
\begin{subfigure}[t]{0.32\textwidth}
  \centering
  \includegraphics[width=\textwidth]{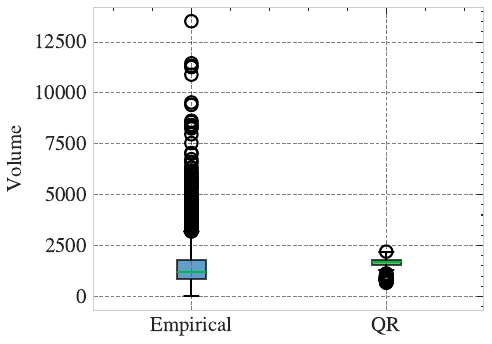}
  \caption{Hourly traded volume.}
\end{subfigure}

\vspace{6pt}

\begin{subfigure}[t]{0.32\textwidth}
  \centering
  \includegraphics[width=\textwidth]{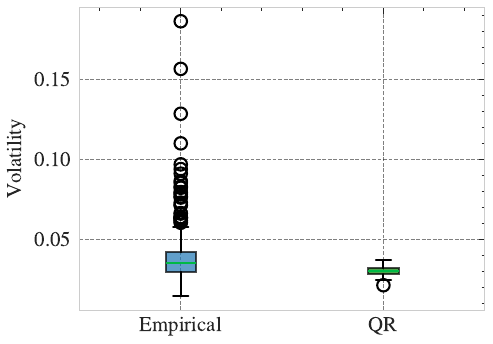}
  \caption{5-min realized volatility.}
\end{subfigure}\hfill
\begin{subfigure}[t]{0.32\textwidth}
  \centering
  \includegraphics[width=\textwidth]{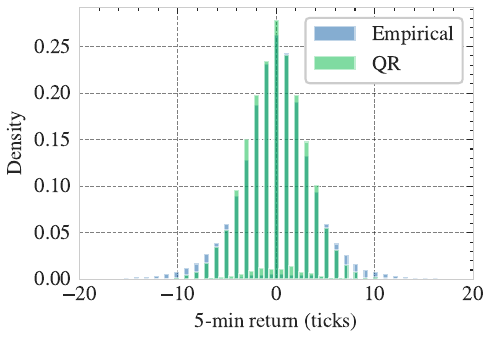}
  \caption{5-min returns distribution.}
\end{subfigure}\hfill
\begin{subfigure}[t]{0.32\textwidth}
  \centering
  \includegraphics[width=\textwidth]{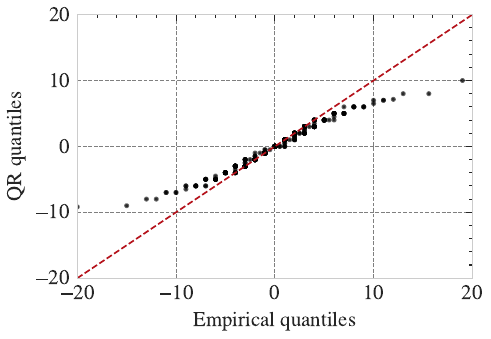}
  \caption{5-min returns QQ plot.}
\end{subfigure}
\tcblower
\caption{Baseline validation statistics. Empirical data (blue) vs.\ QR simulation (green/orange) for VZ.}
\label{fig:validation_VZ}
\end{tcolorbox}
\end{figure}

\begin{figure}[H]
\begin{tcolorbox}[
  enhanced,
  colback={rgb,255:red,235;green,245;blue,251},
  colframe={rgb,255:red,0;green,102;blue,204},
  arc=6pt, boxrule=0.8pt,
  left=4pt, right=4pt, top=4pt, bottom=4pt
]
\centering
\begin{subfigure}[t]{0.32\textwidth}
  \centering
  \includegraphics[width=\textwidth]{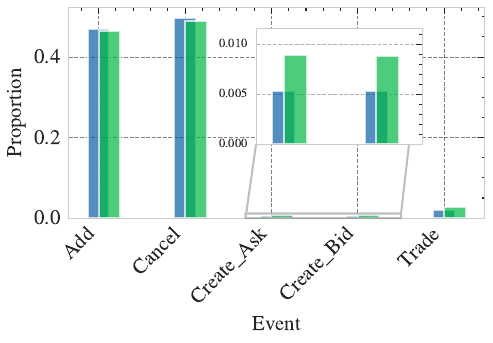}
  \caption{Event-type distribution.}
\end{subfigure}\hfill
\begin{subfigure}[t]{0.32\textwidth}
  \centering
  \includegraphics[width=\textwidth]{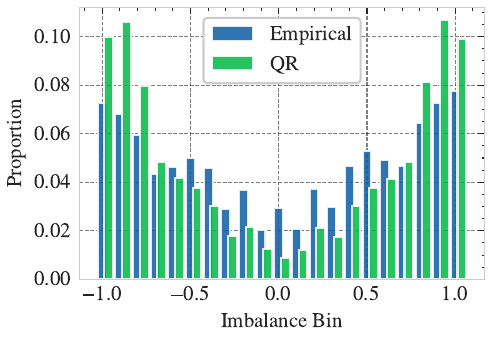}
  \caption{Imbalance before trades.}
\end{subfigure}\hfill
\begin{subfigure}[t]{0.32\textwidth}
  \centering
  \includegraphics[width=\textwidth]{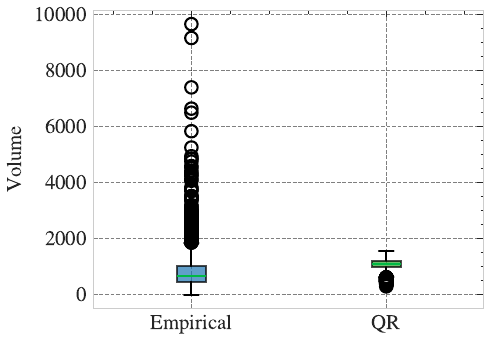}
  \caption{Hourly traded volume.}
\end{subfigure}

\vspace{6pt}

\begin{subfigure}[t]{0.32\textwidth}
  \centering
  \includegraphics[width=\textwidth]{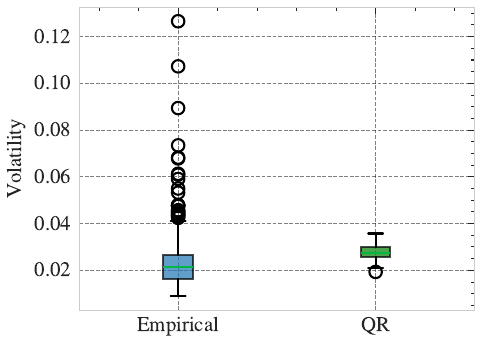}
  \caption{5-min realized volatility.}
\end{subfigure}\hfill
\begin{subfigure}[t]{0.32\textwidth}
  \centering
  \includegraphics[width=\textwidth]{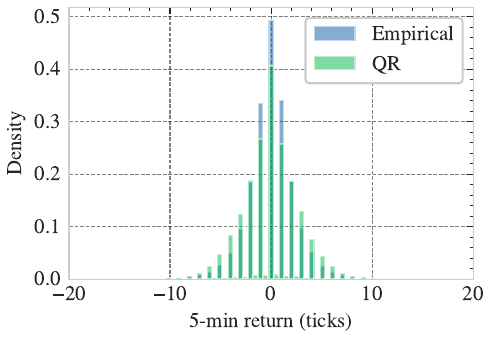}
  \caption{5-min returns distribution.}
\end{subfigure}\hfill
\begin{subfigure}[t]{0.32\textwidth}
  \centering
  \includegraphics[width=\textwidth]{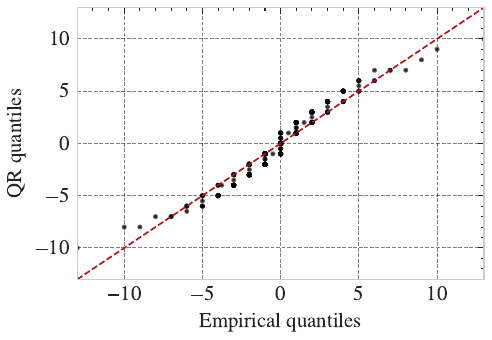}
  \caption{5-min returns QQ plot.}
\end{subfigure}
\tcblower
\caption{Baseline validation statistics. Empirical data (blue) vs.\ QR simulation (green/orange) for T.}
\label{fig:validation_T}
\end{tcolorbox}
\end{figure}

\section{Kernel parameters}
\label{app:kernel_params}

We approximate the target kernel $G(t) = (1 + t/\tau)^{-\beta}$ as a
sum of $K = 12$ exponentials with geometrically-spaced half-lives:
\begin{equation*}
  G(t) \approx \sum_{i=1}^{K} w_i\, e^{-\lambda_i t}, \qquad
  \lambda_i = \frac{\log 2}{h_i},
\end{equation*}
where the half-lives $h_i$ are logarithmically spaced from $0.01$\,s
to $1000$\,s. The weights $w_i \geq 0$ are obtained by non-negative
least squares: we solve
\begin{equation*}
  \min_{w \geq 0} \; \|Xw - y\|^2,
\end{equation*}
where $y_j = G(t_j)$ and $X_{ji} = e^{-\lambda_i t_j}$ on a
logarithmically-spaced time grid $\{t_j\}$. Negative weights are
clipped to zero. The key computational advantage of this
representation is that each component $\phi^{(i)}_t = \sum_{k:\,t_k <
t} w_i\, e^{-\lambda_i(t-t_k)}\,\varepsilon_k\sqrt{V_k}$ satisfies a
simple recursion updated in $O(1)$ per event, so the total cost of
maintaining $\phi_t$ is $O(K)$ per event, independent of the number
of past trades.

Table~\ref{tab:kernel} reports the fitted weights for $\tau = 50$\,s
and $\beta = 1.5$. Most components receive zero weight; the bulk of
the kernel mass concentrates at the 15\,s and 43\,s half-lives, with
a moderate tail contribution out to 1000\,s.

\begin{table}[H]
\centering
\footnotesize
\begin{tabular}{r cccccccccccc}
\toprule
$i$ & 1 & 2 & 3 & 4 & 5 & 6 & 7 & 8 & 9 & 10 & 11 & 12 \\
\midrule
$h_i$ (s) & 0.010 & 0.029 & 0.081 & 0.231 & 0.657 & 1.874 & 5.33 & 15.2 & 43.3 & 123 & 351 & 1000 \\
$w_i$     & 0 & $1.9{\times}10^{-4}$ & 0 & $6.5{\times}10^{-4}$ & 0 & 0 & 0.039 & 0.398 & 0.394 & 0.124 & 0.036 & 0.009 \\
\bottomrule
\end{tabular}
\caption{Sum-of-exponentials kernel weights for $\tau = 50$\,s,
$\beta = 1.5$, $K = 12$. Components 1, 3, 5, 6 receive zero weight.
The bulk of the mass sits at the 15\,s and 43\,s half-lives.}
\label{tab:kernel}
\end{table}

The impact multiplier $m = 0.035$ was calibrated by binary search over
50{,}000 simulated metaorders (10\% of hourly volume, TWAP over
10\,min, each child order of size 2, observed for 60\,min).

\section{Maximum likelihood impact calibration}
\label{app:mle_impact}

An alternative to the simulation-based calibration of
Section~\ref{sec:impact} is to treat $\beta$ as a free parameter and
fit $(\tau, \beta, m)$ jointly by maximum likelihood directly on
observed trade data. The kernel is
$G(t) = (1 + t/\tau)^{-\beta}$
with $\beta > 0$ free, approximated as a sum of exponentials as
described in Appendix~\ref{app:kernel_params}.

\paragraph{Likelihood.}
The bias $b_n = m\,\phi_{t_n}$ modifies the baseline event
probabilities at each step as in~\eqref{eq:bias}. The modified
probability of event $e_n$ in state $\Phi_n$ is
\begin{equation*}
  \tilde{p}_{e_n} = \frac{p_0^{e_n}(\Phi_n) \cdot f_{e_n}(b_n)}{Z_n(b_n)}
\end{equation*}
where $f_{e_n}(b_n) = e^{b_n}$ if $e_n$ is a bid trade and $b_n > 0$,
$f_{e_n}(b_n) = e^{-b_n}$ if $e_n$ is an ask trade and $b_n < 0$,
and $f_{e_n}(b_n) = 1$ otherwise. The normalisation is
\begin{equation*}
  Z_n(b_n) = P_{\text{bid}}(\Phi_n)\,e^{[b_n]^+}
            + P_{\text{ask}}(\Phi_n)\,e^{[-b_n]^+}
            + P_{\text{rest}}(\Phi_n)
\end{equation*}
where $P_{\text{bid}}, P_{\text{ask}}, P_{\text{rest}}$ are the total
baseline probabilities of bid trades, ask trades, and all other events
in state $\Phi_n$. Since $\log p_0^{e_n}$ does not depend on
$(\tau, \beta, m)$, we maximise the reduced log-likelihood:
\begin{equation*}
  \mathcal{L}^*(\tau, \beta, m) = \sum_{n=1}^{N}
    \bigl[\log f_{e_n}(b_n) - \log Z_n(b_n)\bigr].
\end{equation*}
The normalisation term $Z_n > 1$ whenever $b_n \neq 0$, penalising
excessive bias and ensuring a finite optimum.

\paragraph{Optimisation.}
For fixed $(\tau, \beta)$, the kernel weights $\{w_i\}$ are computed
via NNLS and $\phi_{t_n}$ is updated recursively in $O(K)$ per event.
The profile likelihood $\max_m \mathcal{L}^*(\tau, \beta, m)$ is then
evaluated on a coarse grid over $(\tau, \beta)$, with 1D optimisation
over $m$ at each grid point.

\begin{figure}[H]
\begin{tcolorbox}[
  enhanced,
  colback={rgb,255:red,235;green,245;blue,251},
  colframe={rgb,255:red,0;green,102;blue,204},
  arc=6pt, boxrule=0.8pt,
  left=4pt, right=4pt, top=4pt, bottom=4pt
]
\centering
\begin{subfigure}[t]{0.48\textwidth}
  \centering
  \includegraphics[width=\textwidth]{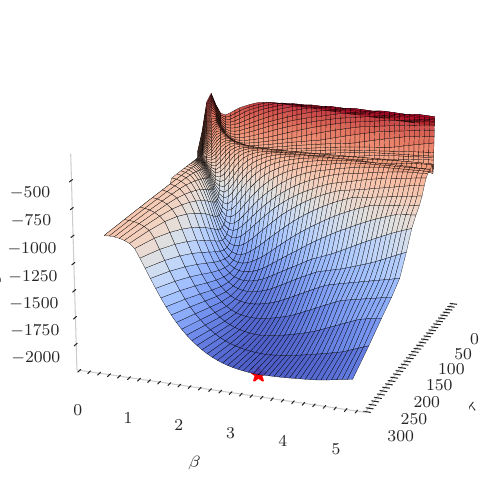}
  \caption{Negative log-likelihood surface over $(\tau, \beta)$ with
    $m$ profiled out.}
  \label{fig:impact_nll}
\end{subfigure}\hfill
\begin{subfigure}[t]{0.48\textwidth}
  \centering
  \includegraphics[width=\textwidth]{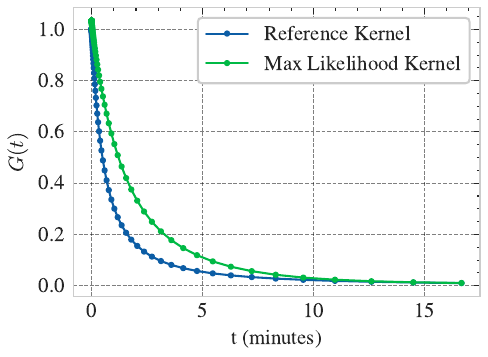}
  \caption{Reference kernel (blue, $\tau=50$\,s, $\beta=1.5$) vs.\
    MLE kernel (green).}
  \label{fig:nll_vs_ref_kernel}
\end{subfigure}
\tcblower
\caption{Maximum likelihood impact calibration results.}
\label{fig:mle_impact}
\end{tcolorbox}
\end{figure}

\paragraph{Results.}
Figure~\ref{fig:impact_nll} (left) shows the negative log-likelihood
surface over $(\tau, \beta)$ with $m$ profiled out. The surface has a
clear, well-identified minimum (red star), confirming that the data are
informative about the kernel shape.
Figure~\ref{fig:nll_vs_ref_kernel} (right) compares the MLE kernel to
the reference kernel of Section~\ref{sec:impact}. Both share the same
qualitative shape, rapid initial decay followed by a power-law tail,
but the MLE kernel decays more slowly over the first few minutes,
reflecting a longer memory of past order flow.

\section{Gaussian mixture model for inter-event times}
\label{app:gmm}

The empirical distribution of $\log_{10} \Delta t$ could in principle
be used directly as a lookup table. We opt instead for a parametric
fit, as the distribution exhibits a visually striking multi-modal
structure in $\log_{10}$ space that is well captured by a Gaussian
mixture. This choice is not essential to the approach --- the
empirical distribution would work equally well --- but it yields a
smooth, compact representation and simplifies sampling.

\begin{figure}[H]
  \centering
  \includegraphics[width=0.48\columnwidth]{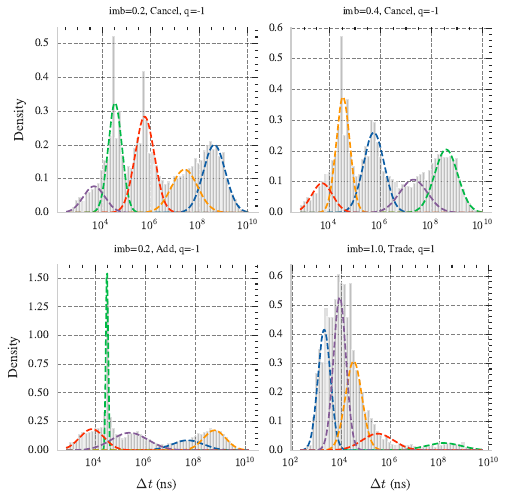}\hfill
  \includegraphics[width=0.48\columnwidth]{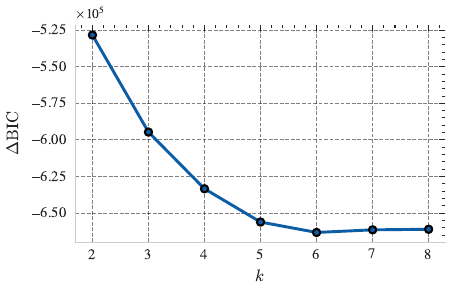}
  \caption{\textbf{Left:} Examples of the fitted 5-component GMM (dashed)
  overlaid on the empirical $\log_{10}\Delta t$ distribution (grey) for
  four $(\Phi, e)$ cells.
  \textbf{Right:} Aggregate $\Delta\mathrm{BIC}(k) = \mathrm{BIC}(k) -
  \mathrm{BIC}(1)$ summed over all $(\Phi, e)$ cells. The elbow at
  $k = 5$ motivates our choice of five mixture components.}
  \label{fig:gmm_examples}
\end{figure}

We model the conditional distribution of inter-event times in
$\log_{10}$ space as a mixture of Gaussians:
\begin{equation*}
  \log_{10} \Delta t \mid \Phi, e \;\sim\;
  \sum_{j=1}^{k} \pi_j \;\mathcal{N}(\mu_j, \sigma_j^2)
\end{equation*}
where the parameters $(\pi_j, \mu_j, \sigma_j^2)$ are estimated by
maximum likelihood via the EM algorithm, independently for each
$(\Phi(\text{LOB}), e)$ cell. Figure~\ref{fig:gmm_examples} shows
examples of the fitted mixtures overlaid on the empirical distributions;
the fit is accurate across a range of imbalance bins and event types.

To select the number of components $k$, we use the Bayesian Information
Criterion (BIC), which balances goodness of fit against model complexity
by penalizing the number of parameters. We sum the BIC across all
$(\Phi, e)$ cells and compare to the single-component baseline
(Figure~\ref{fig:gmm_examples}, right). The aggregate $\Delta\text{BIC}$
decreases sharply up to $k = 5$ and flattens thereafter, so we fix
$k = 5$ globally.

\section{Exchange latency structure across tickers}
\label{app:latency_tickers}

\begin{figure}[H]
    \centering
    \includegraphics[width=0.32\columnwidth]{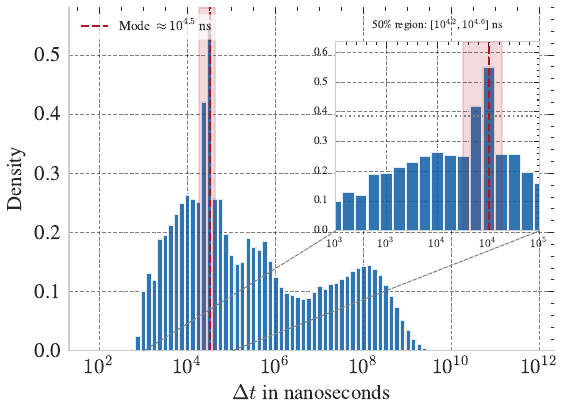}\hfill
    \includegraphics[width=0.32\columnwidth]{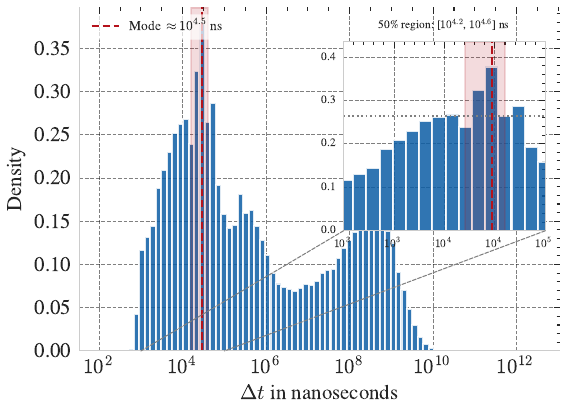}\hfill
    \includegraphics[width=0.32\columnwidth]{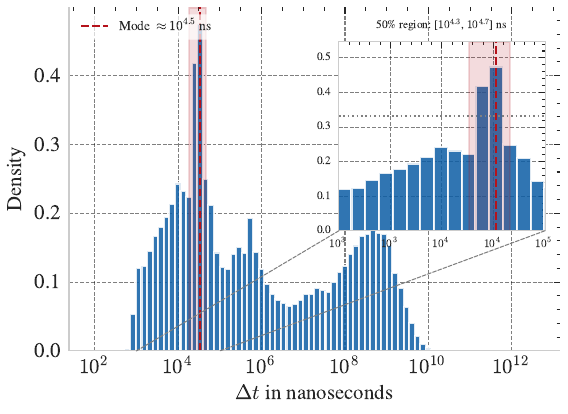}
    \caption{Inter-event time distribution zoomed into the latency mode
    for INTC (left), VZ (centre), and T (right).}
    \label{fig:zoomed_dt_tickers}
\end{figure}

Figure~\ref{fig:zoomed_dt_tickers} shows the inter-event time
distribution zoomed into the latency mode for INTC, VZ, and T. The mode
at $\log_{10}(\Delta t) \approx 4.47$ is consistent across all tickers,
confirming that it reflects exchange-level latency rather than
asset-specific dynamics.

\begin{figure}[H]
    \centering
    \includegraphics[width=0.32\columnwidth]{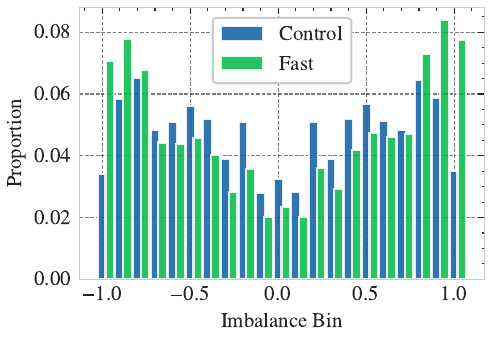}\hfill
    \includegraphics[width=0.32\columnwidth]{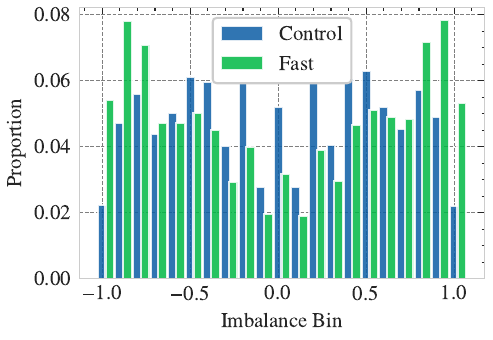}\hfill
    \includegraphics[width=0.32\columnwidth]{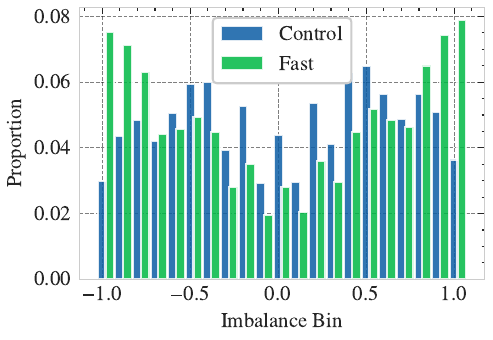}
    \caption{Imbalance before trades: control vs.\ fast events for
    INTC (left), VZ (centre), and T (right).}
    \label{fig:imb_fast_tickers}
\end{figure}

Figure~\ref{fig:imb_fast_tickers} shows the distribution of imbalance
before trades split by inter-event time for INTC, VZ, and T. As with
PFE, fast events ($\Delta t \approx \delta$) are concentrated at extreme
imbalances, supporting the latency race interpretation.

\end{document}